\documentstyle[12pt]{article}
\catcode`\@=11
\def\draftlabel#1{{\@bsphack\if@filesw {\let\thepage\relax
   \xdef\@gtempa{\write\@auxout{\string
      \newlabel{#1}{{\@currentlabel}{\thepage}}}}}\@gtempa
   \if@nobreak \ifvmode\nobreak\fi\fi\fi\@esphack}
        \gdef\@eqnlabel{#1}}
\def\@eqnlabel{}
\def\@vacuum{}
\def\draftmarginnote#1{\marginpar{\raggedright\scriptsize\tt#1}}
\def\draft{\oddsidemargin -.5truein
        \def\@oddfoot{\sl preliminary draft \hfil
        \rm\thepage\hfil\sl\today\quad\militarytime}
        \let\@evenfoot\@oddfoot \overfullrule 3pt
        \let\label=\draftlabel
        \let\marginnote=\draftmarginnote
@
 \def\@eqnnum{(\theequation)\rlap{\kern\marginparsep\tt\@eqnlabel}%
\global\let\@eqnlabel\@vacuum}  }
\def\numberbysection{\@addtoreset{equation}{section}
        \def\theequation{\thesection.\arabic{equation}}}
\def\underline#1{\relax\ifmmode\@@underline#1\else
        $\@@underline{\hbox{#1}}$\relax\fi}
\catcode`@=12
\relax
\numberbysection

\newcommand{\re}{{\rm e}}
\newcommand{\rj}{{\rm j}}
\newcommand{\rjt}{{\rm j}^{\dagger}}
\newcommand{\rJ}{{\rm J}}
\newcommand{\rJt}{{\rm J}^{\dagger}}
\newcommand{\rJo}{{\rm J}_{0}}
\newcommand{\rJot}{{\rm J}_{0}^{\dagger}}
\newcommand{\St}{S^{\dagger}}

\newcommand{\bI}{{\bf I}}
\newcommand{\bJ}{{\bf J}}
\newcommand{\bJo}{{\bf J}_0}
\newcommand{\bJot}{{\bf J}_0^\dagger}
\newcommand{\bJt}{{\bf J}^{\dagger}}
\newcommand{\bg}{{\bf g}}
\newcommand{\bt}{{\bf t}}
\newcommand{\bR}{{\bf R}}
\newcommand{\bRo}{{\bf R}_{0}}

\newcommand{\bS}{{\bf S}}
\newcommand{\bs}{{\bf s}}
\newcommand{\bv}{{\bf v}}
\newcommand{\cA}{{\cal A}}
\newcommand{\cB}{{\cal B}}
\newcommand{\cD}{{\cal D}}
\newcommand{\cE}{{\cal E}}
\newcommand{\cF}{{\cal F}}
\newcommand{\cG}{{\cal G}}
\newcommand{\cH}{{\cal H}}
\newcommand{\cI}{{\cal I}}
\newcommand{\cJ}{{\cal J}}
\newcommand{\cK}{{\cal K}}
\newcommand{\cN}{{\cal N}}
\newcommand{\cO}{{\cal O}}
\newcommand{\cP}{{\cal P}}
\newcommand{\cQ}{{\cal Q}}
\newcommand{\cR}{{\cal R}}
\newcommand{\cT}{{\cal T}}
\newcommand{\cW}{{\cal W}}
\newcommand{\hbI}{\hat{\bf I}}
\newcommand{\hbIo}{\hat{\bf I}_0}
\newcommand{\hIo}{\hat{I}_0}
\newcommand{\hIi}{\hat{I}_1}
\newcommand{\hk}{\hat{k}}
\newcommand{\hp}{\hat{p}}
\newcommand{\hP}{\hat{P}}

\newcommand{\sff}{{\sf f}}

\newcommand{\tk}{\tilde{k}}
\newcommand{\tL}{\tilde{L}}
\newcommand{\un}{^{(n)}}
\newcommand{\uo}{^{(0)}}

\newcommand{\Xol}{{\bf X}_0^{(l)}}
\newcommand{\Xil}{{\bf X}_1^{(l)}}
\newcommand{\an}{\alpha}
\newcommand{\bn}{\beta}
\newcommand{\gn}{\gamma}

\newcommand{\ad}{_{\alpha}}

\newcommand{\aid}{_{\alpha,i}}
\newcommand{\ajd}{_{\alpha,j}}
\newcommand{\adl}{_{\alpha,l}}
\newcommand{\bjd}{_{\beta,j}}

\newcommand{\bd}{_{\beta}}

\newcommand{\abd}{_{\alpha\beta}}
\newcommand{\bad}{_{\beta\alpha}}
\newcommand{\au}{^{(\alpha)}}
\newcommand{\bu}{^{(\beta)}}

\newcommand{\aju}{^{(\alpha,j)}}

\newcommand{\Y}{\Upsilon}
\newcommand{\anum}{\alpha=1,2,3,\quad j=1,2,...,n_{\alpha}}
\newcommand{\bnum}{\beta=1,2,3,\quad j=1,2,...,n_{\beta}}

\newcommand{\Om}{{\Omega}}

\newcommand{\Omt}{\Omega^\dagger}

\newcommand{\Phis}{\Phi^{*}}
\newcommand{\Psis}{\Psi^{*}}
\newcommand{\Rt}{{{\bf R}^3}}
\newcommand{\Rs}{{{\bf R}^6}}

\newcommand{\bC}{{\bf C}}
\newcommand{\Ct}{{{\bf C}^3}}
\newcommand{\Othree}{{\cal O}({\bf C}^3)}
\newcommand{\Opthree}{{\cal O}'({\bf C}^3)}
\newcommand{\Osix}{{\cal O}({\bf C}^6)}
\newcommand{\Opsix}{{\cal O}'({\bf C}^6)}
\newcommand{\Pilh}{\Pi_l^{(\rm hol)}}
\newcommand{\Pibj}{\Pi_b^{(\beta,j)}}
\newcommand{\Sum}{\displaystyle\sum\limits}
\newcommand{\Int}{\displaystyle\int\limits}
\newcommand{\Min}[1]{\mathop{{\rm min}}\limits_{#1}}
\newcommand{\Max}[1]{\mathop{{\rm max}}\limits_{#1}}
\newcommand{\opla}{\mathop{\oplus}\limits}
\newcommand{\Times}{\mathop{\mbox{\large$\times$}}\limits}
\newcommand{\Bigcup}{\mathop{\bigcup}\limits}
\newcommand{\Bigcap}{\mathop{\bigcap}\limits}
\newcommand{\reduction}[2]{\left.\phantom{\bigl|} #1 \right|_{#2}}

\newcommand{\diag}{\mathop{\rm diag}}
\newcommand{\Img}{\mathop{\rm Im}}
\newcommand{\Real}{\mathop{\rm Re}}
\newcommand{\Equal}[1]{\mathop{=}\limits_{#1}}
\newcommand{\be}{\begin{equation}}
\newcommand{\ee}{\end{equation}}

\newcommand{\Frac}[2]{{\displaystyle\frac{#1}{#2}}}
\newtheorem{theorem}{\bf Theorem}
\newtheorem{lemma}{\bf Lemma}

\newtheorem{note}{\bf Remark}
\begin{document}
\thispagestyle{empty}
\large
\begin{center}
{\bf Bogoliubov Laboratory of Theoretical Physics\\[1.5mm] }
{\bf JOINT INSTITUTE FOR NUCLEAR RESEARCH\\[1.5mm]  }
{\bf 141980 Dubna (Moscow region), Russia}
\end{center} 
\vskip-3mm
\hrule{\hfill} 
\vskip 4.5cm
\normalsize

\hfill {Preprint JINR E5--95--45} 

\hfill LANL E-print {\tt nucl-th/9505028}
\bigskip

\bigskip

{\large
\begin{center}
    REPRESENTATIONS FOR THREE-BODY $T$--MATRIX \\[1.5mm]       
             ON UNPHYSICAL SHEETS%
\footnote{
\normalsize Published in Russian 
          in {\em Teoreticheskaya i Matematicheskaya 
          Fizika} {\bf 107} (1996) 450--477 [English translation 
          in {\em Theor. Math. Phys.}] under the title 
          ``Representations for three-body T-matrix 
          on unphysical sheets. I'' }%
\footnote{
\normalsize The work supported in part by Academy of Natural Sciences 
             of RAS and International Science Foundation
             (Grant~\#~RFB000)}%

\end{center}
}
\medskip

{\large
\centerline{  A.K.Motovilov\footnote{
\normalsize  E--mail: MOTOVILV@THSUN1.JINR.DUBNA.SU}  }
}
\bigskip

\bigskip

\bigskip

\normalsize
\centerline{\bf Abstract}
\bigskip

Explicit representations are formulated for the Faddeev 
components of three-body $T$-mat\-rix continued analytically on 
unphysical sheets of the energy Riemann surface.  According to 
the representations, the $T$--mat\-rix on unphysical sheets is 
obviously expressed in terms of its components taken on the 
physical sheet only.  The representations for $T$--mat\-rix are 
used then to construct similar representations for analytical 
continuation of three-body scattering matrices and resolvent. 
Domains on unphysical sheets are described where the 
representations obtained can be applied.

\newpage
\normalsize
\setcounter{footnote}1
\setcounter{page}1
\section{\hspace*{-1em}. INTRODUCTION}
\label{SIntro}
Resonances are one of the most interesting phenomena in 
scattering processes.  The problem of defining and studying 
resonances in quantum mechanics is payed a lot of attention in 
physical and mathematical literature.  In recent years, the 
investigations of resonances in few-particle systems attract a 
growing attention. The role of such resonances is well known in 
physics of nuclear reactions and astrophysics.

Developing methods for studying resonances has a long history 
beginning from the paper by G.Gamow~\cite{GGamow}.  In this 
paper devoted to description of $\alpha$--decay, it was 
discussed for the first time a relation of resonance states to 
complex poles of the scattering matrix (it should be noted 
however that complex frequencies were considered much earlier, 
e.g. by J.J.Thom\-son in 1884).  For spherically symmetrical 
potentials, the interpretation of resonances in two-body 
problems as poles of an analytic continuation of the scattering 
matrix was rigorously based in the known paper by 
R.Jost~\cite{Jost} (for further references in this direction see 
e.g.  the books~\cite{Newton} and~\cite{AlfaroRegge}).

Approaches to interpretation of solutions to the Schr\"odinger 
equation (so--called Gamow's vectors) corresponding to 
resonances are discussed in 
Refs.~\cite{ParGorSud}--\cite{BohmQM} (see also literature cited 
therein). 

Idea to interpret resonances as poles of analytical continuation of 
the resolvent kernel for the Schr\"odinger operator 
(or matrix elements of the resolvent between suitable states) is realized 
in~\cite{Schwinger}--\cite{Hunziker}  (see also Refs. cited in these papers 
and in the books~\cite{Albeverio}, \cite{ReedSimon}).  Such interpretation 
became a basis for the perturbation theory for two--body resonances which 
is well developed now (see.~\cite{Howland}, \cite{Rauch}, \cite{Albeverio}).

In the case when support of interaction in a system of two particles 
is compact with respect to relative coordinate, 
the approach~\cite{LaxP} by P.Lax and R.Phillips may be applied 
(this approach was created initially for acoustical problems). 
The Lax--Phillips approach allows to describe resonances as 
a discrete spectrum of a dissipative operator representing generator 
of contracting semigroup. At present, the Lax--Phillips 
scheme is realized only in those scattering problems which generate
the energy Riemann surface\footnote{
The latter is understood usually as the Riemann surface 
of the resolvent kernel considered as a function of energy or as 
that of the resolvent bilinear form restricted on certain subsets of 
Hilbert space. Such operator--valued functions as the $T$-- and scattering 
matrices have usually the same Riemann surface since these functions 
are closely related to the resolvent.    }
with two sheets of the complex energy plane 
(see Refs.~\cite{PavlovFedorov}, \cite{Fedorov}).
In multichannel scattering problems, the approach above is partly 
realized in~\cite{MotLaxP}.

Beginning from 1970--es, the complex scaling 
method~\cite{BalslevCombes}, \cite{BalslevSkibsted} is applied to 
investigation of resonances (see also Refs.~\cite{Reson} 
and~\cite{ReedSimon}). This method gives a possibility to rotate the 
continuous spectrum of Hamiltonians in such a way that certain 
sectors become accessible for observation on unphysical sheets 
neighboring with physical one. Resonances situated in these sectors 
turn into a part of the discrete spectrum of the Hamiltonian 
transformed.  The complex scaling method may be applied in the cases 
when potentials are analytical functions of coordinates. This method 
allows to compute location of resonances in concrete physical 
problems (see, e.g. Refs.~\cite{Reson}, \cite{Hu}). As regards the  
structure of the scattering matrix and resolvent continued on 
unphysical sheets, this method gives not too large capacities. 

Many important conceptual and constructive results 
(see~\cite{Faddeev63}--\cite{MF}) for the physical sheet in 
three--body scattering problem are known to have been obtained 
on the base of the Faddeev equations~\cite{Faddeev63} and their 
modifications.  In particular the structure of resolvent and 
scattering operator was studied in details, completeness of the 
wave operators was proved and the coordinate asymptotics were 
studied in the case of quickly decreasing as well as Coulomb 
interactions\footnote{
The new approaches~\cite{Enss}, \cite{DerezinskiNbody} 
(see also literature cited in~ \cite{DerezinskiNbody})  have 
been developed recently in abstract scattering theory for 
$N$--body systems which allow to prove existence and asymptotical 
completeness
of the wave operators in the case of pair interactions decreasing 
at the infinity as  $r^{-\varrho}$,
$\varrho>\sqrt{3}-1$, i.e. substantially slower 
than Coulomb potentials.}~\cite{Faddeev63}, \cite{MerkDiss}, \cite{MF},
\cite{EChAYa}.  Analogous results were obtained also for singular 
interactions described by the boundary conditions of various 
types~\cite{EChAYa}, \cite{JMPh}.  On the base of the Faddeev equations, 
the methods of investigation of concrete physical systems were 
developed~\cite{MF}, \cite{EChAYa}, \cite{Schmid}, \cite{Belyaev}.

As to the unphysical sheets, the situation is rather different.  
Here, when solving a concrete $N$--particle problem one usually 
restricts himself with developing some approximate numerical 
algorithm to search for resonances on unphysical sheets 
neighboring with physical one. A survey of different physical 
approaches to study of three-body resonances in the problems of 
nuclear physics can be found in Ref.~\cite{Orlov}.  A number of 
rigorous results (see~\cite{ReedSimon}) is obtained in framework 
of the complex scaling method~\cite{BalslevCombes}, 
\cite{BalslevSkibsted}, \cite{Reson}. These results touch first 
of all the proofs of the existence of analytical continuation of 
resolvent in the $N$--body problem with potentials holomorphic 
with respect to  the scale transforms.  In 
Ref.~\cite{Derezinski}, a proof is given for the existence of 
analytical continuation for the amplitudes of processes 
$2\rightarrow 2$ in the $N$--particle system across the branches 
of continuous spectrum below the first breakup threshold of the 
system into three clusters. 

A goal of the present work consists in analytical continuation 
and investigation of the structure of three--body $T$-matrix,  scattering 
matrices and resolvent 
on unphysical sheets of the energy Riemann surface. 
The interaction potentials are supposed to be pairwise 
and decreasing in coordinate space not slower than exponentially. 
When constructing a theory of resonances in the two--body problem 
with such interactions one can use the coordinate as well as momentum 
representations. However, it is clear { a priori} that the analytical 
continuation of the  three--body  scattering theory 
equations~\cite{Faddeev63}, \cite{MF} on unphysical sheets  
becomes a very difficult problem if the equations are written 
in configuration space. Thing is that there exist noncompact (cylindrical) 
domains where interactions do not decrease. Meanwhile, the kernels of the 
integral equations continued increase exponentially. 
Their solutions have to increase exponentially, too. 
This means that the integral terms become divergent ones and the 
coordinate space equations lose a sense. 
In the momentum space, the integral terms of the scattering 
theory equations, e.g. the Faddeev equations for components of 
$T$--matrix, are actually the Cauchy type integrals analytical 
continuation of which (in a sense of distributions) is a solvable 
problem. A continuation of such kind on unphysical sheets 
neighboring with physical one was already realized for the s--wave 
Faddeev equations in the paper~\cite{OrlovTur} (see also 
Ref.~\cite{Orlov}) for the case of separable pair potentials. 
In the present paper, we construct a continuation of the Faddeev 
equations in the case of sufficiently arbitrary pair potentials 
not only on the neighboring unphysical sheets but also on all those 
remote sheets 
of the three--body Riemann surface where is possible to guide 
the spectral parameter (the energy $z$) going around two--body 
thresholds.  

Main result of the paper consists in a basing of existence of analytical 
continuation on unphysical sheets of $z$ for the Faddeev components 
$M\abd(z)$,\,\, $\an,\bn=1,2,3,$ of the operator $T(z)$ 
and a construction of representations for them 
in terms of the physical sheet [see formula (\ref{Ml3fin})].  
According to the representations, the continued matrix   $M(z)$ 
of the Faddeev components, $M=\{ M\abd\}$, is explicitly expressed 
on unphysical sheets in terms of 
this matrix itself taken on the physical one and some truncations 
of the scattering matrix. Kind of the truncation is determined  by 
the index (number) of the unphysical sheet concerned. 
Note that structure of the representations 
is quite analogous to that of the representations 
found in the author's recent works~\cite{MotTMF} and~\cite{MotYaF} 
for analytical continuation of $T$--matrix in multichannel scattering 
problems with binary channels. Representations for analytical continuation 
of three--body scattering matrices follow immediately from the 
representation above for $M(z)$
[see Eqs.~(\ref{Slfin}) and~(\ref{R3l})].
As follows from the explicit representations  (\ref{Ml3fin}), (\ref{Slfin})
and (\ref{R3l}) obtained by us, 
the singularities of $T$--matrix, scattering matrices 
and resolvent on unphysical sheets differing
from those on the physical one (poles at the discrete spectrum 
eigenvalues of the Hamiltonian),  are actually singularities 
of the operator--valued functions of $z$ inverse with respect 
to suitable truncations of the scattering matrix. 
Consequently, the resonances 
(i.e. the poles of $T$--matrix, scattering matrix 
and resolvent on unphysical sheets)  are zeros of certain truncations 
of the scattering matrix taken on the physical sheet.

Results of the present paper were announced in the 
report~\cite{MotFewBodyCEBAF}.

The paper is organized as follows.

In Sec.~\ref{SNotations}, the main notations are described. 
Sec.~\ref{St2body} contains an information on analytical properties of 
the two--body $T$-- and scattering matrices 
which is necessary in subsequent sections. 
Sec.~\ref{SMSphys} is devoted to description of properties 
of the Faddeev components of three--body $T$--matrix and scattering 
matrices on the physical sheet of energy. In particular, 
the domains  on the physical sheet are established where the 
half--on--shell Faddeev components and different truncations of 
the scattering matrices included in the 
representations (\ref{Ml3fin}), (\ref{Slfin}) and (\ref{R3l})
may be considered as holomorphic functions. 
We justify these representations only on 
a certain part of the three--body Riemann surface which is described 
in Sec.~\ref{SRiemannSurface}.
Analytical continuation of the Faddeev equations on unphysical 
sheets is described in Sec.~\ref{SRepresent}. Also, in this section,  
the representations  
(\ref{Ml3fin}), (\ref{Slfin}) are (\ref{R3l}) formulated 
for analytical continuation of the matrix  $M(z)$, scattering matrices 
and resolvent respectively. 
\section{\hspace*{-1em}. NOTATIONS}
\label{SNotations}
We consider a system of three spinless non--relativistic quantum 
particles. Movement of the mass center is assumed to be separated. 
For description of the system we use standard sets of the relative momenta  
$k\ad$, $p\ad$ \cite{MF}.
For example 
\begin{equation}
\label{Jacobi}
\begin{array}{rcl}
k_1 & = &
\left[\Frac{{\rm m}_2+{\rm m}_3}{2{\rm m}_2 {\rm m}_3}\right]^{1/2}
\cdot
\Frac{{\rm m}_2 {\rm p}_3 -{\rm m}_3 {\rm p}_2}
{{\rm m}_2+{\rm m}_3}   \\
p_1 & = &
\left[    \Frac{{\rm m}_1 +{\rm m}_2 +{\rm m}_3 }
{ 2{\rm m}_1 ({\rm m}_2 + {\rm m}_3) }    \right]^{1/2}\cdot
\Frac{({\rm m}_2+{\rm m}_3){\rm p}_1-{\rm m}_1 ({\rm p}_2 + {\rm p}_3)}
{ {\rm m}_1 +{\rm m}_2 +{\rm m}_3 } ,
\end{array}
\end{equation}
where ${\rm m}\ad$, ${\rm p}\ad$
are masses and momenta of particles. 
Expressions for  $k\ad$, $p\ad$ with $\an=2,3$
may be obtained from  (\ref{Jacobi})
by cyclic permutation of indices.
Usually we combine relative momenta  $k\ad$, $p\ad$
into six-vectors $P= \lbrace k\ad, p\ad \rbrace$.
A choice of certain pair  $\lbrace k\ad, p\ad \rbrace$ fixes 
cartesian coordinate system in ${\bf R}^{6}$.
Transition from one pair of momenta to another one means rotation in 
${\bf R}^{6}$,
$ k\ad=c\abd k\bd+s\abd p\bd$,\,\,
$ p\ad=-s\abd k\bd +c\abd p\bd,$\,\,
with coefficients $c\abd$, $s\abd$
depending on the particle masses only \cite{MF}, such that  
$ -1< c\abd < 0,$ $s\abd^2=1-c\abd^2 $,   
$c\bad=c\abd$ and $s\bad=-s\abd$, \,\, $\bn\neq\an$.

In momentum representation, the Hamiltonian $H$
of the three--body system under consideration is given by 
$(Hf)(P)=P^{2}f(P)+\sum_{\an=1}^{3} (v\ad f)(P),$
$P^2=k\ad^2+p\ad^2$, $f\in \cH_0\equiv L_2(\Rs)$,
with $v\ad$, the pair potentials which are integral operators 
in $k\ad$ with kernels  $v\ad(k\ad, k'\ad)$.

For the sake of definiteness all the potentials 
$v\ad$,\,\, $\an=1,2,3,$ are supposed to be local.
This means that the kernel of  $v\ad$
depends on the difference of variables $k\ad$ and $k'\ad$ only,
$v\ad(k\ad, k'\ad)$ $=v\ad(k\ad - k'\ad)$.
We consider two variants of the potentials $v\ad$.
In the first one,   $v\ad (k)$ are holomorphic functions of the variable 
$k\in\Ct$ which satisfy the estimate 
\be
\label{vpot}
|v\ad (k)|\leq \Frac{c}{(1+|k|)^{\theta_{0}}}
{\rm e}^{a_{0}|{\rm Im} k|} \quad \forall k\in\bC^3
\ee
with some  $c>0,$ $a_0 >0$
and $\theta_{0}\in (3/2,\, 2)$.\,\,
In the second variant, the potentials $v\ad (k)$
are holomorphic functions with respect to   $k$
in the strip $W_{2b} =\{ k: \, k\in\Ct,\,
|{\rm Im} k|<2b \}$
only and obey for $k\in W_{2b}$
the condition (\ref{vpot})
with $a_0 =0$:
\be
\label{vpotb}
|v\ad (k)|\leq \Frac{c}{(1+|k|)^{\theta_{0}}}
 \quad \forall k: |\Img k|< 2b.
\ee
It is supposed that in both variants $v\ad (-k)=$ $\overline{v\ad (k)}$.
The latter condition guarantees self--adjointness of the Hamiltonian
$H$ on the set 
$\cD(H)=\lbrace f:\, \int (1+P^{2})^{2}|f(P)|^{2} dP <\infty \rbrace$
\cite{Faddeev63}.

Note that the first variant requirements 
of holomorphness of $v\ad (k)$ in all  $\Ct$
and no more than exponential increasing (\ref{vpot})
in $|{\rm Im}\, k|$
mean that these potentials have a compact support in the coordinate space. 
In the second variant, the potentials $v\ad (k)$
rewritten in the coordinate representation, decrease 
exponentially.

By $h\ad$, $(h\ad f)(k\ad)=$ $k\ad ^{2} f(k\ad)+(v\ad f)(k\ad)$,
we denote the Hamiltonian of the pair subsystem $\an$.  The operator $h\ad$
acts in $L_2 (\Rt)$.  Due to conditions (\ref{vpot}) and (\ref{vpotb}) 
its discrete spectrum  $\sigma_{d}(h\ad)$ 
is negative and finite~\cite{ReedSimon}.
We enumerate the eigenvalues  $\lambda\ajd\in\sigma_{d}(h\ad),$
$\lambda\ajd < 0$, $ j = 1,2,...,n\ad$, $n\ad<\infty$, 
taking into account their multiplicity: 
number of times to meet an eigenvalue in the numeration equals 
to its multiplicity. Maximal of these numbers is denoted by 
$\lambda_{\rm max}$,\,
$\lambda_{\rm max}=\Max{\an,j}\lambda\ajd<0.$
Notation $\psi\ajd(k\ad)$ is used for respective 
eigenfunctions.

By $\sigma_d(H)$ and $\sigma_c(H)$ we denote respectively the 
discrete and continuous components of the spectrum 
$\sigma(H)$ of the Hamiltonian 
$H$. Note that   $\sigma_c(H)=(\lambda_{\rm min},+\infty)$ with
$\lambda_{\rm min}= \Min{\an ,j} \lambda\ajd$.

Notation $H_0$ is used for the operator of kinetic energy, 
$(H_0 f)(P)=$ $P^{2}f(P)$.  $R_0 (z)$ and $R(z)$ stand for the resolvents 
of the operators 
$H_0$ and $H$ : $R_0 (z)=(H_0
-zI)^{-1}$ and $R(z)=(H -zI)^{-1}$ where in this case, $I$ is 
the identity operator in  $\cH_0$.

Let 
$ M\abd (z) = \delta\abd v\ad -v\ad R(z) v\bd$,
$\an, \bn =1,2,3,$
be the Faddeev components~\cite{Faddeev63}, \cite{MF}
of the three--body  $T$--matrix 
$ T(z)=V-VR(z)V $ with $V=v_1 +v_1 +v_3 $.
Operators $ M\abd (z) $
satisfy the Faddeev equations~\cite{Faddeev63}, \cite{MF}
\be
\label{Fadin}
M\abd (z) = \delta\abd \bt\ad(z) -
\bt\ad(z) R_0 (z)\Sum_{\gn\neq\an} M_{\gamma\bn}(z), 
\quad \an=1,2,3.
\ee
Here, the operator ${\bt}\ad(z)$ has the kernel 
\be
\label{tpair3}
{\bt}\ad(P,P',z)=t\ad(k\ad,k'\ad,z-p\ad^2)
\delta(p\ad-p'\ad),
\ee
where $t\ad(k,k',z)$
stands for the kernel of the pair  $T$--matrix 
$t\ad(z)=v\ad-v\ad r\ad(z)v\ad$ with 
$r\ad(z)=(h\ad-z)^{-1}$.

It is convenient to rewrite the system (\ref{Fadin}) 
in the matrix form 
\be
\label{MFE}
M(z)=\bt (z) - \bt (z) \bRo (z) \Y M(z),
\ee
with $\bt (z)=\diag\{\bt_1(z),\bt_2(z),\bt_3(z)\}$ and 
$\bRo (z)=\diag\{R_0(z),R_0(z),R_0(z)\}$. By $\Y$
we denote a number $3\!\times\! 3$--matrix with the elements 
$\Y\abd=1-\delta\abd.$
$ M(z) $ is the operator matrix 
constructed of the components
$ M\abd (z) $, $ M=\{ M\abd \},$ $\an,\bn=1,2,3$. The matrices  
$M$, $\bt$, $\bRo$
and $\Y$ are considered as operators in the Hilbert space 
${\cal G}_0=\opla_{\an=1}^{3} L_2(\Rs)$.
By $ \cQ^{(k)}(z),$ \,\,
$
 \cQ^{(k)}(z)=\bigl( - \bt(z)\bRo(z)\Y \bigr)^k \bt(z),
$
we denote iterations of the absolute term $\cQ\uo(z)=\bt(z)$
of (\ref{MFE}).

The resolvent $R(z)$ is expressed in terms of the matrix $M(z)$
by formula~\cite{MF}
\be
\label{RMR}
R(z)=R_0(z)-R_0(z) \Om M(z) \Omt R_0(z),
\ee
where $\Om$, $\Om: \cG_0\rightarrow\cH_0$, stands for 
the matrix--row, $\Om=(1,\,\, 1,\,\, 1)$. At the same time 
$\Omt=\Om^{*}=(1,\,\, 1,\,\, 1)^{\dagger}$.
The symbol ``$\dagger$'' means transposition.

Throughout the paper we understand by
$\sqrt{z-\lambda}$,\,\,\, $z\!\in\!\bC$,\,\,
$\lambda\!\in\!\bR,$\,  the main branch of the function 
$(z-\lambda)^{1/2}$.  By $\hat{q}$ we denote usually the unit vector 
in the direction  $q\!\in\!\bR^N$,\,\,
$\hat{q}={q}/{|q|}$, and by $S^{N-1}$ the unit sphere 
in $\bR^N$,\,\, $\hat{q}\!\in\!  S^{N-1}$.  The inner product in 
$\bR^N$ is denoted by $(\,\cdot\, ,\,\cdot\,)$.  Notation  
$\langle\,\cdot\, ,\,\cdot\, \rangle $ is used for 
inner products in Hilbert spaces. 

Let $\cH^{\aju}=L_2(\Rt)$
and $\cH^{\au}=\opla_{j=1}^{n\ad} \cH^{\aju}$.
By $\Psi\ad$
we denote operator acting from  ${\cal H}^{\au}$
to ${\cal H}_0$
as 
$
(\Psi\ad f)(P)=\Sum_{j=1}^{n\ad} \psi\ajd (k\ad)f_j(p\ad),
$\,\,\,
$ f=( f_1, f_2, ... , f_{n\ad} )^{\dag}$.
Notation $\Psi\ad^{*}$
is used for the operator adjoint to  $\Psi\ad$.
By $\Psi$ we denote the block--diagonal matrix operator 
$\Psi =\diag\{\Psi_1,$ $\Psi_2,$ $\Psi_3\}$
which acts from  ${\cal H}_1 =\opla_{\an=1}^{3} {\cal H}\au$
to ${\cal G}_0$ and by 
$\Psi^{*}$, 
the operator adjoint to $\Psi$.
Analogously to $\Psi\ad$, $\Psi\ad^{*}$, $\Psi$
and  $\Psi^{*}$
we introduce the operators $\Phi\ad$, $\Phi\ad^{*}$, $\Phi$
and  $\Phi^{*}$,
which are obtained from the former by replacement of the eigenfunctions 
$\psi\ajd(k\ad)$
with the form--factors $\phi\ajd(k\ad)=(v\ad\psi\ajd)(k\ad)$,
$\an=1,2,3$, $j=1,2,...,n\ad$.

The pair  $T$--matrix  $t\ad(z)$ is known~\cite{Faddeev63},\cite{MF} 
to be an analytical operator--valued function of the variable 
$z\in{\bf C}\backslash[0,+\infty)$ 
having at the points $z\in\sigma_d(h\ad)$
simple poles. Its kernel admits the representation 
\be
\label{tdistr}
t\ad(k,k',z)= - \Sum_{j=1}^{n\ad} \Frac{\phi\ajd(k)
\overline{\phi\ajd}(k')}{\lambda\ajd-z}
+\tilde{t}\ad(k,k',z),
\ee
where $\tilde{t}\ad(k,k'z)$
is a function holomorphic in  $z\in{\bf C}\backslash[0,+\infty)$.
Therefore 
\be
\label{tcd}
\bt\ad(z)=-\Phi\ad\bg\ad(z)\Phi\ad^{*}+\tilde{\bt}\ad(z),
\ee
where the operator $\tilde{\bt}\ad(z)$ has the kernel 
$\tilde{t}\ad(k\ad,k'\ad,z-p\ad^{2})\delta(p\ad-p'\ad)$
and  $\bg\ad(z)$,
$\bg\ad(z)$=$\diag\{ g_{\alpha,1}(z),$ $...,$ $g_{\an,n\ad}(z)\}$, 
is the block--diagonal matrix with elements  $g\ajd(z)$, 
the operators in ${\cal H}^{(\an,j)}$
with singular kernels 
$
g\ajd(z)(p\ad,p'\ad,z)=
{\delta(p\ad-p'\ad)}
\left/{\bigl(\lambda\ajd-z+p\ad^2\bigr)}\right. .
$

Below, we consider restrictions of different functions 
on the energy shell
\be
\label{EnergyS-twobody}
   k=\sqrt{z} \hat{k},\quad  \hat{k}\in S^2,
\ee
in the two--body problem and on the 
energy shells 
\be
\label{EnergyS3}
   P=\sqrt{z} \hat{P},\quad  \hat{P}\in S^5,
\ee
and 
\be
\label{EnergyS2}
  p\ad=\sqrt{z-\lambda\ajd} \hat{p}\ajd,\quad
    \hat{p}\ajd\in S^2, \quad \anum,
\ee
in the problem of three particles. 
In the last case the sets  (\ref{EnergyS3}) and (\ref{EnergyS2}) 
are called respectively three--body and two--body energy shells. 
	
Let $ {\cal O}({\bf C}^{N})$
be the Fourier transform of the space $ C_{0}^{\infty}(\bR^{N})$  
(we deal with  $N=3$ or $N=6$ only). Any  $f(q)\in {\cal O}({\bf C}^{N})$
is a holomorphic function in variable 
$q=(q_1,$ $ q_2,...,$ $q_N) \in {\bf C}^{N} $
satisfying the estimates 
$
\left| \frac{\partial^{|m|}}
{\partial q^{m_1}_{1}\cdot...\cdot\partial q^{m_N}_{N}}
f(q) \right| \leq c_{\theta}(f)\cdot
{\exp(a|\Img q|)}{(1+|q|)^{-\theta}},
$
with $a$, the radius of the ball centered in the origin and containing 
the support of the Fourier pre--image of this function 
in ${\bf R}^N $,
$
\quad |m|=m_1+...+m_N ,
$
and
$
|\Img q | = \sqrt{\sum_{j=1}^{N} |\Img q_j|^2 }.
$
As $\theta$ one can take arbitrary positive number.
For fixed $f$ and $ m=(m_1,...,m_N)$, the coefficient $ c_\theta > 0 $
depends only on $ \theta $.

Let $\rj(z)$ be the operator restricting  
functions $f(k)$, $k\in \Rt,$
on the energy shell (\ref{EnergyS-twobody}) at 
$z=E\pm i0$,\,\, $E>0$,
and continuing them if possible, on the domain 
of complex values of the energy  $z$.
On the set ${\cal O}(\Ct)$ the operator  $\rj(z)$ acts as 
\be
\label{j2}
\rj(z)f(\hat{k})=f(\sqrt{z}\hat{k}).
\ee
Its kernel is the holomorphic generalized function 
(distribution)~\cite{GShilov}
$\rj(\hat{k},k',z)=\delta(\sqrt{z}\hat{k}-k')$.

By $\rjt(z)$ we denote the operator 
``transposed'' with respect to  $\rj(z)$.
Acting on $\varphi\in L_2(S^2)$ the operator $\rjt(z)$ gives as a result 
the generalized functions (distributions) over  ${\cal O}(\Ct),$
\be
\label{j2t}
(\rjt(z)\varphi)(k)=
\Int_{S^2} d\hat{k}\,\,\delta(k-\sqrt{z}\hat{k'})\,\,\varphi(\hat{k})=
\Frac{\delta(|k|-\sqrt{z})}{z} \varphi(\hat{k}),
\ee
i.e. 
\be
\label{j2tint}
(\rjt(z)\varphi,f)=\Int_{S^2}d\hat{k}f(\sqrt{z}\hat{k})\varphi(\hat{k}),
\quad f\in\Othree.
\ee
Remember that in terms of the operators  $\rj(z)$
and  $\rjt(z)$, 
the pair scattering matrices  $s\ad(z)$, \,\,
$s\ad(z): L_2(S^2)\rightarrow L_2(S^2)$,
look as (index of pair is omitted)~\cite{MotTMF}:
\be
\label{s2}
s(z)=\hat{I} + {\rm a}_0 (z) \rj(z)t(z)\rjt(z),
\ee
where ${\rm a}_0(z) = -\pi i \sqrt{z}$
and  $\hat{I}$ is the identity operator in  $L_2 (S^2)$.
Analyticity domain of $s(z)$, $z\in\bC$, is determined in a general way 
by properties of the pair potential  $v$
(see e.g.,~\cite{Newton}, \cite{AlfaroRegge},
and also~\cite{MotTMF}, ~\cite{MotYaF}).

Let $\rJ\ajd(z)$, \,\, $\anum$, be the  operator
of restriction on the energy shell 
(\ref{EnergyS2}). Its action on  $\Othree$ is defined as 
$$
(\rJ\ajd(z)f)(\hat{p}\ad)=f(\sqrt{z-\lambda\ajd}\,\,\hat{p}\ad),
\quad \an=1,2,3, \quad j=1,2,...,n\ad.
$$
Operators $\rJ\ajd(z)$ have the kernels 
$
\rJ\ajd(\hat{p}\ad,p'\ad,z)=\delta(\sqrt{z-\lambda\ajd}\hat{p}\ad-p'\ad).
$

By $\rJo(z)$ we denote operator of restriction 
on the shell  (\ref{EnergyS3}). On $ \Osix $
this operator is defined as  
$
(\rJo(z)f)(\hat{P})=f(\sqrt{z}\hat{P}).
$
Its kernel is 
$
\rJo(\hat{P},P',z)=(\sqrt{z})^{-5}\delta(\sqrt{z}\hat{P}-P')=
{\delta(\sqrt{z}-|P'|)}
\delta(\hat{P},\hat{P'}).
$

Notations $\rJt\ajd(z)$ and $\rJot(z)$
are used for respective ``transposed''
operators. Their action is defined similarly to (\ref{j2t}),(\ref{j2tint})
as 
$$
(\rJt\ajd(z)\varphi)(p\ad)=
\Int_{S^2}d\hat{p}'\ad
\delta(p\ad-\sqrt{z-\lambda\ajd}\hat{p}'\ad)
\varphi(\hat{p}'\ad),\quad \varphi\in \hat{\cH}\aju,
$$
$$
(\rJot(z)\varphi)(P)=
\Int_{S^5}d\hat{P}'
\delta(P-\sqrt{z}\hat{P}')
\varphi(\hat{P}'),\quad \varphi\in \hat{\cH}_0,
$$
where $\hat{\cH}\aju\equiv L_2(S^2)$ and $\hat{\cH}_0\equiv L_2(S^5).$
The generalized functions $\rJt\ajd(z)\varphi$
and $\rJot(z)\varphi$ are elements of the spaces  $\Opthree$
and $\Opsix $ of distributions over $\Othree $ and $\Osix $, respectively.

Operators $\rJ\ajd$ and $\rJt\ajd$ are then combined into the block--diagonal 
matrices
$\rJ\au(z)=$ $\diag\{ \rJ_{\alpha,1}(z),$ $...,\rJ_{\alpha,n\ad}(z) \} $
and $\rJ^{(\alpha)\dagger}(z)=$ $
\diag\{ \rJt_{\alpha,1}(z),...,$ $\rJt_{\alpha,n\ad}(z) \}. $
Latter are used to construct operators 
$\rJ_1 (z)=$ $\diag\{ \rJ^{(1)}(z),$ $\rJ^{(2)}(z),$
$\rJ^{(3)}(z) \}$
and $\rJt_1 (z)=$ $\diag\{ \rJ^{(1)\dagger}(z),$ $\rJ^{(2)\dagger}(z),$
$\rJ^{(3)\dagger}(z) \}.$
The action of $\rJ\au(z)$ and $\rJ_1(z)$ on elements 
of the spaces respectively, 
$\cO\au=$ $\Times_{\an=1}^{n\ad} \cO\aju$,
$\cO\aju\equiv\Othree$ and $\cO_1=\Times_{\an=1}^3 \cO\au$ can be 
understood 
by the definition of the operators $\rJ\ajd(z)$.
The operators  $\rJ^{(\alpha)\dagger}(z)$
act from $\hat{\cH}\au\equiv\opla_{j=1}^{n\ad} \hat{\cH}\aju $
to the space of analytical distributions  $\cO\au{}'$
over $\cO\au$.
In its turn the operator $\rJt_1(z)$ acts from  
$\hat{\cH}_1=\opla_{\an=1}^{3}\hat{\cH}\aju$
to the space of analytical distributions $\cO'_1$ over $\cO_1$.

At last, we use the block--diagonal operator 
$3\!\times\! 3$--matrices 
$\bJo(z)=\diag\{ \rJo(z),\rJo(z),\rJo(z) \} $
and $\bJot(z)=\diag\{ \rJot(z),\rJot(z),\rJot(z) \}, $
constructed of the operators $\rJo(z)$ and $\rJot(z)$, respectively 
as well as the operators $\bJ(z)=\diag\{ \rJo(z),\rJ_1(z) \} $ 
$\bJt(z)=\diag\{ \rJot(z),\rJt_1(z) \}. $
Action of these operators is clear due to definitions 
of the operators $\rJo,$ $\rJ_1$, $\rJot$ and $\rJt_1$.
In particular the operator  $\bJt(z)$  acts from  the space 
$\hat{\cG}_0=\opla_{\an=1}^3 \hat{\cH}_0$ to the space 
$\Times_{\an=1}^3 \cO'(\bC^6)$.

The identity operators in the spaces 
$\hat{\cH}_0$,  $\hat{\cG}_0$,  $\hat{\cH}_1$ and 
$\hat{\cH}_0\oplus\hat{\cH}_1$ are denoted by 
$\hIo$, $\hbIo$, $\hIi$ and $\hbI$ respectively.

\section{\hspace*{-1em}. ANALYTICAL CONTINUATION OF THE $T$-- AND SCATTERING 
\newline
                         MATRICES IN THE TWO--BODY PROBLEM}
\label{St2body}

In this section we remember some analytical properties of the pair 
$T$--matrices which will be necessary further when posing the three--body 
problem. Note that above properties are well known 
(see e.g., Refs~\cite{AlfaroRegge}, \cite{Newton} and also~\cite{Orlov}) 
for a wide class of the potentials  $v\ad(x)$.  As a matter of fact we 
want to expose here only an explicit representation for the two--body 
$T$--matrix on unphysical sheet which is a particular case of 
the explicit representations constructed in the author's work~\cite{MotTMF} 
(see Theorem~2 in~\cite{MotTMF} and comments to it) 
for a rather more general situation of analytical continuation 
of $T$--matrix on unphysical sheets in the multichannel problem 
with binary channels. 

Throughout the section we shall consider a fixed pair subsystem. 
Therefore its index will be omitted in notations. Statements will 
be given for the first variant of the potentials (\ref{vpot}). 
If it will be necessary, different  assertions for the second 
variant (\ref{vpotb}) will be written in brackets. 
Also, we use the notation 
\be
\label{Pib}
\cP_b=\left\{ z:\,\,\, \Real z>-b^2+\Frac{1}{4b^2}(\Img z)^2 \right\}.
\ee

Remember that the energy Riemann surface in the two--body problem 
coincides with that of the function $z^{1/2}$.  On the physical sheet,  
$z^{1/2}=\sqrt{z}$, and on the unphysical one, $z^{1/2}=-\sqrt{z}$.
For these sheets we use the notations respectively, $\Pi_0$ and $\Pi_1$.

Representation for continuation of $t(z)$ on unphysical sheet 
which will be used further, is described by the following statement 
which is one--channel variant of Theorem~2 of Ref.~\cite{MotTMF}.
\begin{theorem}\label{Tht2body}\hspace*{-0.5em}{\sc .}
The two--body $T$--matrix $t(z)$ allows analytical continuation 
in variable  $z$ on the sheet $\Pi_1$
(on the domain  $\cP_b\bigcap\Pi_1$) as a bounded operator in $L_2(\Rt)$.
Result of the continuation 
$\reduction{t(z)}{\Pi_1}$ $\left(\reduction{t(z)}{\cP_b\bigcap\Pi_1}\right)$
is expressed by $T$-- and $S$--matrices on the physical sheet:
\be
\label{3tnf}
\reduction{t(z)}{\Pi_1}=t(z)-{\rm a}_0(z)\,\, \tau(z)
\ee
where $\tau(z)=(t\rjt s^{-1}\rj t)(z)$.
The kernel $\reduction{t(k,k',z)}{\Pi_1}$
is a holomorphic function of variables 
$k,k'\in \Ct $ and 
$z\in\Pi_1\setminus\bigl(\sigma_{\rm res}\bigcup\sigma_d(h)\bigr)$
\,\,\,\,
$\biggl( k,k'\in W_b$ and 
$
z\in\cP_b\bigcap\Pi_1\setminus
\bigl(\sigma_{\rm res}\bigcup\sigma_d(h) \bigr)\,\,\biggr)$.
Here, $\sigma_{\rm res}$ is a set of the points 
$z\in\bC\setminus\overline{\sigma(h)}$
$\left(z\in\cP_b\setminus\overline{\sigma(h)}\right)$
where the operator $[s(z)]^{-1}$ does not exist.
\end{theorem}
Emphasize that for the second variant  of potentials 
(\ref{vpotb}), the existence of the continuation of  $t(z)$ on 
unphysical sheet is guaranteed by Theorem~\ref{Tht2body} for the 
domain  $\cP_b\bigcap\Pi_1$
bounded by the parabola $\Img\sqrt{z}=b$, inside of which 
the function $v\left(\sqrt{z}(\hk-\hk')\right)$ is holomorphic in 
$z$ for arbitrary $\hk,\hk'\in S^2$.
Note also that the operator $\bigl(\rj t\rjt\bigr)(z)$,
included in Eq.~(\ref{s2}), is a compact operator in $C(S^2)$~\cite{MotTMF}.
Consequently on the domain of its analyticity 
$\Pi_0\setminus\overline{\sigma(h)}$
\,\,\, $\bigl(\,\cP_b\bigcap\Pi_0\setminus\overline{\sigma(h)}\,\bigl)$
on the physical sheet,  one can apply to the equation 
\be
\label{zero2}
s(z){\cal A}=0
\ee
the Fredholm alternative~\cite{ReedSimon} (see Ref.~\cite{MotTMF}).
This means that the set $\sigma_{\rm res}$ being countable, 
has not concentration points in $\bC\setminus\overline{\sigma(h)}$
$\left(\cP_b\setminus\overline{\sigma(h)}\right)$.

On the physical sheet $\Pi_0$, 
the pair  $T$--matrix 
admits the representation (\ref{tdistr}).
It follows from the Lippmann--Schwinger equation for  
$\phi_j$, $j=1,2,...,n,$
\be
\label{phij}
\phi_j(k)=-\Int_\Rt dq\,\, v(k,q)\Frac{1}{q^2-\lambda_j}\phi_j(q),
\quad \lambda_j < 0,
\ee
that form--factor $\phi_j(k)$
admits analytical continuation in $k$
on  $\Ct$ (on $W_{2b}$)  and at the same time, it satisfies the type 
(\ref{vpot}) estimate where one has to replace $\theta_0$
with a number $\theta$, $1 < \theta < \theta_0 $,
which can be taken in any close vicinity of $\theta_0$~\cite{Faddeev63}.
Hence the eigenfunction 
\be
\label{psiphij}
\psi_j(k)=-\Frac{\phi_j(k)}{k^2-\lambda_j}
\ee
of $h$ admits also an analytical continuation 
on $\Ct$ (on $W_{2b}$)
with the exception of the set $\{ k\in\Ct:k^2=\lambda_j\}$
where $\psi_j(k)$
has singularities (turning for $k=\sqrt{z}\hat{k}$,
$\hat{k}\in S^2$,
into a pole in energy $z$ at $z=\lambda_j$).

The regular summand $\tilde{t}(k,k'z)$ of the kernel of $t(z)$
is holomorphic function in variables $k,k'\in\Ct$, $z\in\Pi_0$ \,\,\,
($k,k'\in W_b$, $z\in \cP_b\bigcap\Pi_0$)
and admits the estimate 
$$
|\tilde{t}(k,k'z)|<{c}{(1+|k-k'|)^{-\theta} }\cdot
\exp[a(|\Img k|+|\Img k'|)],
$$
with arbitrary $\theta\in(1,\theta_0)$.

As to continuation of $\reduction{t(z)}{\Pi_1}$,
it follows from Eq.~(\ref{3tnf})
that the points $z\!\in\!\sigma_d(h)$
give to it generally speaking, poles of the first order.
One can easily check however that if eigenvalue 
$\lambda\in\sigma_d(h)$ is simple then the 
respective singularities of the both summands of  (\ref{3tnf})
compensate each other and the pole of $\reduction{t(z)}{\Pi_1}$
does not appear at  $z=\lambda$.
It follows from the Fredholm analytical alternative~\cite{ReedSimon}
for Eq.~(\ref{zero2}) only that 
poles of $\reduction{t(z)}{\Pi_1}$
at $z\in \sigma_{\rm res}$
are of a finite order and no more.
It is easily to show that if 
$\cA(\hk)$ is a nontrivial solution of Eq.~(\ref{zero2}) at 
$z\in\sigma_{\rm res}$,
$z\not\in\sigma_d(h)$,
then the Schr\"odinger equation  
has at this $z$ a nontrivial (resonance) solution. Asymptotics 
of this solution $\psi^{\#}_{\rm res}(x)$ 
in configuration space, $x\in\bR^3$, 
is exponentially increasing: 
$
\psi^{\#}_{\rm res}(x)\Equal{x\rightarrow\infty}
\left({\cal A}(-\hat{x})+o(1)\right)
\Frac{{\rm e}^{-i\sqrt{z}|x|}}{|x|}.
$
The function $\psi^{\#}_{\rm res}(x)$
is so--called Gamow vector corresponding to resonance at the energy $z$
(see e.g., Refs.~\cite{Newton}, \cite{BohmJMP}, \cite{BohmQM}).
The function  ${\cal A}(\hat{k})$
makes a sense to the breakup amplitude of the resonance
state\footnote{Analogous assertion takes place as well 
in the multichannel scattering problem with 
$m$ binary channels: solution 
$
{\cal A}=({\cal A}_1,{\cal A}_2,...,{\cal A}_m)
$
to the equation $s_l(z){\cal A}=0$
at resonance energy $z\in\sigma_{\rm res}^l$
(in notations of Ref.~\cite{MotTMF}) represents amplitudes 
(i.e. coefficients at spherical waves in 
coordinate asymptotics of the channel 
components of solution to respective Schr\"odinger equation) 
$
{\cal A}_1(\hat{k}_1),{\cal A}_2(\hat{k}_2),...,{\cal A}_m(\hat{k}_m)
$
of resonance on the sheet $\Pi_l$ to breakup into channels 
1,2,...,$m$, respectively.}.

The formula for analytical continuation of the scattering matrix 
on unphysical sheet  $\Pi_1$
(on the set $\cP_b\bigcap\Pi_1$)
follows immediately from Eq.~(\ref{3tnf})
(see Ref.~\cite{MotTMF}),
\be
\label{S2P1}
\reduction{s(z)}{\Pi_1}={\cal E}[s(z)]^{-1}{\cal E},
\ee
where ${\cal E}$ stands for the inversion in $L_2(S^2)$,
$({\cal E}f)(\hat{k})=f(-\hat{k})$.

Utilizing (\ref{3tnf}) one can easily to get the explicit representation 
in terms of the physical sheet as well for analytical continuation 
on $\Pi_1$ $\left(\mbox{ on } \cP_b\bigcap\Pi_1 \right)$
of the resolvent $r(z)$ kernel\footnote{
Similar representations take place as well in the case of the multichannel 
problem. In notations of Ref.~\cite{MotTMF}
read them as 
$
\reduction{r(z)}{\Pi_l}=
r+(I-rv)\rJt AL s^{-1}_l \rJ (I-vr).
$
 }:
\be
\label{res2}
\reduction{r(z)}{\Pi_l}=
r+{\rm a}_0(I-rv)\rjt s^{-1} \rj (I-vr).
\ee
The continuation has to be understood in a sense 
of generalized functions (distributions) over  $\Othree$:
one has to continue the bilinear form 
$
\Phi(z)=\bigl(r(z)f_1,f_2\bigr)\equiv
{\displaystyle\Int}_{\Rt} dk {\displaystyle\Int}_{\Rt} dk'\,
f_2(k)\, r(k,k',z) f_1(k')
$
%
\section{\hspace*{-1em}. MATRIX $M(z)$ AND THREE--BODY SCATTERING 
\newline
                         MATRICES ON THE PHYSICAL SHEET}
\label{SMSphys}

At the beginning, remember shortly principal 
properties~\cite{Faddeev63}, \cite{MF} of the Faddeev equations (\ref{MFE})
for the matrix $M(z)$  and properties of the kernels $M\abd(P,P',z)$
at real arguments $P,P'\in\Rs $.
To formulate these properties we cite here 
the following definition~\cite{Faddeev63}.

The operator--valued function $\cQ\abd(z)$
of variable $z\in\bC$,\,\, $\cQ\abd(z):\,\, \cH_0\rightarrow\cH_0,$
is the type $\cD\abd$ function,\,\, $\an,\bn=1,2,3,$
if it admits the representation 
\begin{eqnarray}
\nonumber
\lefteqn{  \cQ\abd(z)=\cF\abd(z)+\Phi\ad\bg\ad(z)\cI\abd(z)+} \\
\label{QDrepr}
 & & +\cJ\abd(z)\bg\bd(z)\Phis\bd+
      \Phi\ad\bg\ad(z)\cK\abd(z)\bg\bd(z)\Phis\bd.
\end{eqnarray}
The operator--valued functions
$\cF\abd(z):\,\, \cH_0\rightarrow\cH_0,$\,\,\,
$\cI\abd(z):\,\, \cH_0\rightarrow\cH\au,$\,\,\,
$\cJ\abd(z):\,\, \cH\bu\rightarrow\cH_0$\,\,
and 
$\cK\abd(z):\,\, \cH\bu\rightarrow\cH\au$\,\,
are called components of the function $\cQ\abd(z)$.
If $\cQ\abd(z)$ is an integral operator then its kernel is called 
kernel of the type $\cD\abd$.

Let 
$
\cN(P,\theta)=\Sum_{\an,\bn,\,\,\an\!\ne\!\bn}
(1+|p\ad|)^{-\theta} (1+|p\bd|)^{-\theta}.
$
A function $\cQ(z)$ of the type  $\cD\abd$
is called the class $ \cD_{\an\bn}(\theta,\mu)$ function if its 
components $\cF\abd$, $\cI\abd$, $\cJ\abd$ and $\cK\abd$
are integral operators and for the kernels  $\cF\abd(P,P',z)$
at $P,$ $P',$ $\Delta P,$ $\Delta P'\in\Rs$, 
the estimates
\be
\label{Est1}
\left| \cF\abd(P,P',z) \right| \leq\, c\,  \cN(P,\theta)
\bigl(1+{p\bd'}^2 \bigr)^{-1},
\ee
\begin{eqnarray}
\nonumber
\lefteqn{ \left|\cF\abd(P+\Delta P,P'+\Delta P',z+\Delta z)-
\cF(P,P',z)\right|\leq}     \\
\label{Est2}
 & & \leq \, c\, \cN(P,\theta) \bigl(1+{p\bd'}^2 \bigr)^{-1}
(|\Delta P|^\mu+|\Delta P'|^\mu+|\Delta z|^\mu)
\end{eqnarray}
with certain $c>0$ take place and at the same time,   
the kernels 
$\cI_{\an,j;\bn}(p\ad,P',z),$\,\,
$\cJ_{\an;\bn,k}(P,p'\bd,z)$
and $\cK_{\an,j;\bn,k}(p\ad,p'\bd,z)$
satisfy inequalities which may be got from (\ref{Est1}) and (\ref{Est2}) 
if to take respectively, $k\ad=0,$ \,\, $k'\bd=0$
or simultaneously, $k\ad=0,$  $k'\bd=0$.

Let $\cQ\un(z)$ be an iteration of the absolute term of Eq.~(\ref{MFE}).
In a contrast to $\cQ^{(0)}(z)=\bt(z)$
kernels of the operators $\cQ\un(z)$ at  $n > 0$
do not include $\delta$--functions. Moreover, it follows from the 
representation (\ref{tcd}) for $\bt\ad(z)$ explicitly manifesting 
a contribution of the discrete spectrum of pair subsystems, that 
matrix elements $\cQ\abd\un(z)$,\,\,
$\an,\bn=1,2,3,$ of the operators $\cQ\un(z)$ with $n\geq 1$
are actually functions of the $\cD\abd$ type.
Their components $\cF\abd\un(z)$, $\cI\abd\un(z)$,
$\cJ\abd\un(z)$ and $\cK\abd\un(z)$
at $z\in\bC\setminus[\lambda_{\rm min},+\infty)$
are bounded operators depending on $z$ analytically.
In the case of potentials (\ref{vpot}) and (\ref{vpotb}), 
the H\"older index of smoothness $\mu$ for their kernels with respect to  
variables $P,P',p\ad$ and $p'\bd$
at $z\not\in[\lambda_{\rm min},+\infty)$
equals to 1. If $n\leq 3$\,\,
then as $\Img z\rightarrow 0,$\,\, $\Real z\in[\lambda_{\rm min},+\infty)$
the kernels $\cF_{\an\bn}\un$ \,\,
$\cI_{\an,j;\bn}\un,$\,\,
$\cJ_{\an;\bn,k}\un$,\,\,
and $\cK_{\an,j;\bn,k}\un$
have so--called {\em minor} (three--particle) singularities 
(see Refs.~\cite{Faddeev63} and~\cite{MF}) 
weakening with growing $n$.
At $n\geq 4$ such singularities do not appear at all 
and these kernels become H\"older functions in all their variables including 
the limit values  $z=E\pm i0,$ \,\, $E\in(\lambda_{\rm min},+\infty).$
More precise statement~\cite{Faddeev63} is following: 
the operator--valued functions $\cQ\abd\un(z)$ at $n\geq 4$
belong to the type $\cD\abd(\theta,\mu)$, \,\, $0<\theta<\theta_0$, \,
$0<\mu<\frac{1}{8}$, uniformly with respect to   $z$ 
changing on arbitrary bounded set in the complex plane 
$\bC$ with cut along the ray $[\lambda_{\rm min},+\infty)$.
One can take as $\theta$, $\theta<\theta_0$, any number 
as close as possible to $\theta_0$.
Thus, instead of $M(z)$ it is convenient~\cite{Faddeev63} 
to come to the new unknown 
$
\cW (z)=$ $ M(z)- $ $ \sum_{n=0}^{3} \cQ\un(z),
$
satisfying the equation 
\be
\label{MFEW}
\cW (z)=\cW \uo(z)-\bt(z)\bRo(z)\Y \cW (z)
\ee
analogous to Eq.~(\ref{MFE}) but with another absolute term 
$\cW \uo(z)=\cQ^{(4)}(z)$.

Immersion of Eq.~(\ref{MFEW}) in the Banach space 
$\cB(\theta,\mu)$ (a description of the latter see in Refs.~\cite{Faddeev63},
or~\cite{MF}) leads one to the following important 
\begin{theorem}\label{ThFaddeev1}{\rm (L.D.Faddeev~\cite{Faddeev63})}
\hspace*{-0.5em}{\sc .}
{Eq.~(\ref{MFE}) is uniquely solvable at 
$z\not\in\overline{\sigma_d(H)}$. Its solution $M(z)$ admits 
the representation  
\be
\label{MQW}
   M(z)=\Sum_{n=0}^{3}\cQ\un(z) +\cW (z),
\ee
where the operator--valued function $\cW (z)$ is holomorphic 
in variable $z$ at $z\not\in\overline{\sigma(H)}$ 
and its components $\cW\abd(z)$
belong to the classes $\cD\abd(\theta,\mu)$,
\,\, $3/2 < \theta < \theta_0$, $0 < \mu < \frac{1}{8}$, \,\,
uniformly with respect to   $z$ changing in  arbitrary bounded set 
of the complex plane 
$\bC$ with cut along the ray $[\lambda_{\rm min},+\infty)$ 
and removed neighborhoods of the points of $\sigma_d(H)$.}
\end{theorem}

Remember now  structure of the scattering operator $\bS $ ~\cite{Faddeev63},
\cite{MF} for the system of three particles. 
For this purpose we introduce the operator--valued function 
$\cT(z)$,\,\, $\cT(z):\,\cH_0\oplus\cH_1\rightarrow\cH_0\oplus\cH_1$,
of $z\in\bC\setminus\overline{\sigma(H)}$,
\be
\label{T3body}
\cT(z)\equiv \left(
\begin{array} {cc}
\Omega M(z)\Omega^{\dagger}   &    \Omega M(z)\Y\Psi  \\
\Psi^{*}\Y M(z)\Omega^{\dagger} &   \Psi^{*}(\Y\bv +\Y M(z)\Y)\Psi
\end{array}\right),
\ee
with $\bv=\diag\{ v_1, v_2, v_3 \}.$
Note that $\cT_{00}(z)=\Om M(z)\Omt\equiv T(z)$, 
$\cT_{00}(z):\cH_0\rightarrow\cH_0$.
The rest of the components $\cT_{01}(z):\cH_1\rightarrow\cH_0$,\,\,
$\cT_{10}(z):\cH_0\rightarrow\cH_1$ and 
$\cT_{11}(z):\cH_1\rightarrow\cH_1$ is expressed by 
the transition operators~\cite{MF} (see also~\cite{Schmid}) \,\,
$U_0(z)=\Om M(z)\Y$,\,\, $U_0^\dagger=\Y M(z)\Omt$
and $U(z)=\Y\bv +\Y M(z) \Y$: \,\,\,
$\cT_{01}=U_0\Psi$,\,\,\, $\cT_{10}=\Psis U_0^\dagger$
and $\cT_{11}=\Psis U\Psi.$  
The operator $\cT(z)$ is a matrix integral operator 
with kernels 
$\cT_{00}(P,P',z)$,\,\,  $\cT_{\an,i;\, 0}(p\ad,P',z)$,\,\,\,
$\cT_{0;\, \bn,j}(P,p'\bd,z)$ and $\cT_{\an,i;\, \bn,j}(p\ad,p'\bd,z)$,
\,\,\, $\an=1,2,3,$ \,\, $i=1,2,...,n\ad,$\,\, $\bnum$,
properties of which are determined including 
the limit points  $z=E\pm i0$,\,\, $E > \lambda_{\rm min}$,
by Theorem~\ref{ThFaddeev1}.

By $\hat{\cT}(z)$,\,\,
$\hat{\cT}(z):$
$\hat{\cH}_0\oplus\hat{\cH}_1\rightarrow\hat{\cH}_0\oplus\hat{\cH}_1$,
we denote analytical continuation 
in $\bC^\pm$
(see Theorems~\ref{ThLTL}, \ref{ThJ0TJ0t} and~\ref{ThJ0MYPsiJ1t})
of the operators $\hat{\cT}(E\pm i0)$ having the kernels 
$$
\begin{array}{lcl}
\bigl(\hat{\cT}(E\pm i0) \bigr)_{00}(\hat{P},\hat{P}') & = &
\cT_{00}(\pm\sqrt{E}\hat{P},\, \pm\sqrt{E}\hat{P}',E\pm i0), \quad E > 0; \\
\bigl(\hat{\cT}(E\pm i0) \bigr)_{0;\,\bn,j}(\hat{P},\hat{p}'\bd) & = &
\cT_{0;\,\bn,j}(\pm\sqrt{E}\hat{P},\,
\pm\sqrt{E-\lambda\bjd}\hat{p}'\bd,E\pm i0),
\quad E > 0; \\
\bigl(\hat{\cT}(E\pm i0) \bigr)_{\an,i;0}(\hat{p}\ad,\hat{P}') & = &
\cT_{\an,i;0}(\pm\sqrt{E-\lambda_{\an,i}}\hat{p}\ad,\,
\pm\sqrt{E}\hat{P}',E\pm i0), \quad
E > 0 ; \\
\bigl(\hat{\cT}(E\pm i0) \bigr)_{\an,i;\,\bn,j}(\hat{p}\ad,\hat{p}'\bd)
& = & \cT_{\an,i;\,\bn,j}(\pm\sqrt{E-\lambda_{\an,i}}\hat{p}\ad,\,
\pm\sqrt{E-\lambda\bjd}\hat{p}'\bd,E\pm i0),   \\
&  & \phantom{E > Max \{ \lambda_{\an,i},\, \lambda\bjd\}}
    E > \Max{}\{ \lambda_{\an,i},\, \lambda\bjd\}.
\end{array}
$$
We assume by definition that the product $\bigl(\bJ\cT\bJt\bigr)(z)$
coincides with $\hat{\cT}(z)$,
\be
\label{mcteq}
(\bJ\cT\bJt)(z)=\left(
\begin{array}{cc}
\bigl(\rJo \cT_{00}\rJot \bigr)(z)  & \bigl(\rJo \cT_{01}\rJt_1 \bigr)(z) \\
\bigl(\rJ_1 \cT_{10}\rJot \bigr)(z)  & \bigl(\rJ_1 \cT_{11}\rJt_1 \bigr)(z)
\end{array}\right)
\equiv \hat{\cT}(z).
\ee
Elements of the matrix $(\bJ\cT\bJt)(z)$
are expressed in terms of amplitudes of different processes taking 
place in the three--body system under consideration~\cite{MF} 
(see also Sec.~7 of~\cite{Mot3Unphys2}). 

The scattering operator $\bS $ is unitary one in the space 
$\cH_0 \oplus \cH_1 $ and as well as $\cT$,
it has a natural block structure. 
Its components 
$\bS_{00}$, $\bS_{0;\beta,j}$, $\bS_{\alpha,i;0}$, $\bS_{\alpha,i;\beta,j}$
have the kernels, respectively  
\begin{eqnarray}
\label{S00}
\bS_{00}(P,P') & = &
\delta(P-P')- 2\pi i\, \delta(P^2-P'^2) \cT_{00}(P,P', P'^2+i0),  \\
\label{S0b}
\bS_{0;\beta,j}(P,p'\bd) & = &
-2 \pi i\, \delta(P^2-p'^{2}\bd-\lambda\bjd)
\cT_{0;\beta,j}(P,p'\bd,\lambda\bjd+p'^{2}\bd+i0),  \\
\label{Sa0}
\bS_{\alpha,i;0}(p\ad, P') & = &
- 2 \pi i\, \delta(\lambda_{\alpha,i}+p^{2}\ad - P'^2)
\cT_{\alpha,i;0}(p\ad, P',P'^2 +i0),                  \\
\label{Sab}
\bS_{\alpha,i;\beta,j}(p\ad,p'\bd) & = &
\delta\abd\delta_{ij}\delta(p\ad-p'\bd)-      \\
& & -2 \pi i\, \delta(\lambda_{\alpha,i} +p\ad^2 -\lambda_{\beta,j} -p\bd^2)
\cT_{\alpha,i;\beta,j}(p\ad,p'\bd,\lambda_{\beta,j} +p\bd^2 + i0).
\nonumber
\end{eqnarray}

Scattering matrices arise from  $\bS$ in the spectral 
decomposition for $H$ as operators acting in the ``cross 
section'' (at fixed energy) of the space ${\cal H}_0\oplus {\cal 
H}_1$ in the Neumann direct integral~\cite{MerkDiss}.  
Extraction of the scattering matrix from $\bS$ is related as a 
matter of fact to the  replacements $|P|^2\rightarrow E$, 
$\lambda_{\alpha,i} +p\ad^2 \rightarrow E$, $\an=1,2,3,$ 
$i=1,2,...,n\ad$, in expressions (\ref{S00})---(\ref{Sab}) and 
then to the  factorization of dependence of the kernels of $\bS$ 
on the energies $E$ and $E'$,
\be
\label{SEE}
\bS(E,E')=-\pi i \delta(E-E')\tilde{\vartheta}(E)S'(E+i0)
\tilde{\vartheta}(E'),
\ee
where $\tilde{\vartheta}(E)$ is a diagonal matrix--function 
constructed of the Heaviside functions 
${\vartheta}(E)$
and ${\vartheta}(E-\lambda\bjd)$: \,
$\tilde{\vartheta}(E)=$
$\diag\{\vartheta(E),\vartheta(E-\lambda_{1,1}),...$
$\vartheta(E-\lambda_{1,n_1}),\vartheta(E-\lambda_{2,1}),...$
$\vartheta(E-\lambda_{2,n_2}),\vartheta(E-\lambda_{3,1}),...$
$\vartheta(E-\lambda_{3,n_3}) \}. $
At $z\in\bC$ we understand by $S'(z)$ 
the operator--valued function 
$
S'(z)=A^{-1}(z)\, \hbI + \hat{\cT}(z).
$
Here and all over further,  $A(z)=\diag\{A_0(z),\, A_1(z)\}$ with 
$A_0(z)=-\pi i z^2$ \,  
and 
$A_1(z)=\diag\{ A^{(1)},$ $A^{(2)},$ $A^{(3)} \}$ where 
in its turn,  
$A\au (z)= \diag\{ A_{\an,1}(z),...,A_{\an,n\ad}(z) \}$
with $A\ajd(z)=-\pi i \sqrt{z-\lambda\ajd}.$

Continuing the factorization,
$
S'(z)=S(z)A^{-1}(z)=A^{-1}(z)S^{\dagger}(z),
$
corresponding to separating in (\ref{SEE}) the multiplier 
$-\pi i A^{-1}(E+i0)$
as a derivative of measure in the Neumann integral
 above~\cite{MerkDiss} for  $\cH_0 \oplus \cH_1 $, one comes to 
the scattering matrices 
\be
\label{SMx}
S(z)= \hbI + \bigl(\bJ\cT\bJt \, A \bigr)(z) \,\mbox{ and }\,
\St(z)= \hbI + \bigl(A\,\bJ\cT\bJt \bigr)(z) .
\ee
In a contrast to Ref.~\cite{MerkDiss} it is more convenient for us 
to use namely this, nonsymmetrical, form of the scattering matrices. 
Matrices $S(z)$ and $\St(z)$
are considered as operators in $\hat{{\cal H}}_0\oplus \hat{{\cal H}}_1 $.
At $z=E+i0$, $E>0$, these operators are unitary. At $z=E+i0$, $E<0$,
there are certain truncations of $S(z)$ and $\St(z)$ 
determined by the number of open channels which are 
unitary in $\hat{{\cal H}}_0\oplus \hat{{\cal H}}_1 $, 
namely the matrices 
$\tilde{S}(E)=
\hat{\bf I}+ \tilde{\vartheta}(E)\bigl(S(E+i0)-\hat{\bf I}\bigr)
\tilde{\vartheta}(E)$
and 
$\tilde{\St}(E)=
\hat{\bf I}+ \tilde{\vartheta}(E)\bigl(\St(E+i0)-\hat{\bf I}\bigr)
\tilde{\vartheta}(E)$.
It follows from Eq.~(\ref{SMx}) that operator $\cT$ may be considered 
as a kind of ``multichannel $T$--matrix'' (cf. Ref.~\cite{MotTMF})
for the system of three particles.

Note that the matrix $\cT(z)$ may be replaced in Eq.~(\ref{SMx})
with the matrix $\cT^\dagger(z)$ obtained from  
$\cT(z)$ by the substitution $\Y\bv\rightarrow\bv\Y$
(respectively, $U \rightarrow U^{\dagger}=\bv\Y+ \Y M \Y$)
in the second component of the lower row of (\ref{T3body}).
To prove that 
$\bigl(\bJ\cT^\dagger\bJt \bigr)(z)=\bigl(\bJ\cT\bJt \bigr)(z)$,
it is sufficient to observe that for 
$z=E\pm i0$, $E>\lambda\ajd,$ $\anum$,
\be
\label{JPvPJ}
(\rJ_1 \Psis\Y\bv\Psi\rJt_1)(z)=
(\rJ_1 \Psis\bv\Y\Psi\rJt_1)(z).
\ee
Indeed, according to Eqs.~(\ref{phij})
and~(\ref{psiphij}),
\be
\label{PYvP}
(\Psis\Y\bv\Psi)_{\an,i;\bn,j}(p\ad,p'\bd)=
-\Frac{1-\delta\abd}{|s\abd|^{3}} \,\cdot\,
\Frac{    \overline{\phi}_{\an,i}(\tk\ad\bu(p\ad,p'\bd))\,\,
\phi\bjd(\tk\bd\au(p'\bd,p\ad))    }
{ [\tk\ad\bu(p\ad,p'\bd)]^2-\lambda_{\an,i}   },
\ee
\be
\label{PvYP}
(\Psis\bv\Y\Psi)_{\an,i;\bn,j}(p\ad,p'\bd)=
-\Frac{1-\delta\bad}{|s\abd|^{3}} \,\cdot\,
\Frac{    \overline{\phi}_{\an,i}(\tk\ad\bu(p\ad,p'\bd))\,\,
\phi\bjd(\tk\bd\au(p'\bd,p\ad))    }
{ [\tk\bd\au(p'\bd,p\ad)]^2-\lambda\bjd   }
\ee
where 
\be
\label{ktilde}
\tk_{\gamma}^{(\delta)}(q,q'))=
\Frac{-c_{\gamma\delta} q+q'}{s_{\gamma\delta} }, \quad
\gamma,\delta=1,2,3,
\ee
$ q,q'\in\Rt $ { (we shall suppose later that $q,q\in \Ct$)}.
One can easily to understand that on the energy shells 
$|p\ad|=\sqrt{E-\lambda_{\an,i}},$ \,\,\,
$|p'\bd|=\sqrt{E-\lambda\bjd},$ \,\,\,
$E>\lambda_{\an,i},$ \,\,\,
$E>\lambda\bjd,$
the denominators of the fractions (\ref{PYvP})
and (\ref{PvYP})
coincide,
\begin{eqnarray}
\nonumber
\lefteqn{ (\tk\ad\bu)^2-\lambda_{\an,i}=(\tk\bd\au)^2-\lambda\bjd= } \\
\label{k2lambda}
& &=\Frac{1}{|s\abd|^2}(E-\lambda_{\an,i}+E-\lambda\bjd-
2c\abd\sqrt{E-\lambda_{\an,i}}\sqrt{E-\lambda\bjd}
(\hp\ad,\hp'\bd)-s\abd^2 E).
\end{eqnarray}
Meanwhile the expression  (\ref{k2lambda})
can not become zero at $E>\lambda_{\an,i},$ $E>\lambda\bjd$
(see Lemma~\ref{LEqQuadr}).
It follows now from Eqs.~(\ref{PYvP}), (\ref{PvYP}) and~(\ref{k2lambda})
that the equality~(\ref{JPvPJ}) is true.

Along with  $S(z)$ and $\St(z)$  we shall consider also 
the {\em truncated} scattering matrices 
\be
\label{Slcut}
 S_l(z) \equiv \hat{\bf I} + (\tL\bJ \cT \bJt L A)(z) \, \mbox{ and }\,
\St_l(z) \equiv \hat{\bf I} + (A L\bJ \cT \bJt\tL)(z),
\ee
where the multi--index 
\be
\label{lmulti}
l=(l_0,l_{1,1},...,l_{1,n_1},l_{2,1},...,l_{2,n_2},l_{3,1},...,l_{3,n_3})
\ee
has the components $l_0=0$ or  $l_0=\pm 1$
and $l\ajd=0$ or $l\ajd=1$, $\anum$.
By $L$ and $\tL$ we denote the diagonal number matrices 
\be
\label{Lmatrix}
L=\diag\{ l_0,l_{1,1},...,l_{1,n_1},l_{2,1},...,
l_{2,n_2},l_{3,1},...,l_{3,n_3}    \}
\ee
and 
\be
\label{Ltmatrix}
\tL=\diag\{ |l_0|,l_{1,1},...,l_{1,n_1},l_{2,1},...,
l_{2,n_2},l_{3,1},...,l_{3,n_3}    \},
\ee
corresponding to the multi--index $l$.
The matrix $\tL$ is evidently to 
be a projector in $\hat{\cH}_0 \oplus \hat{\cH}_1$
on the subspace $\hat{\cH}_1^{(l)}$  if $l_0=0$ or 
on the subspace $\hat{\cH}_0 \oplus \hat{\cH}_1^{(l)}$ if $l_0\neq 0$.
Here in both cases, 
$\hat{\cH}_1^{(l)} =\opla_{l\ajd\neq 0} \hat{\cH}\aju.$

As can be seen from formulas (\ref{SMx}) and (\ref{T3body}) the 
scattering matrices $S(z)$ and $\St(z)$ include kernels 
$M\abd(P,P',z)$ taken on the energy shells: their arguments 
$P\in\Rs$ and $P'\in \Rs$ are connected with the energy $z=E+i0$ 
by Eqs.~(\ref{EnergyS3}) at $E > 0$ or~(\ref{EnergyS2}) at $E > 
\lambda\ajd$. We establish below [see formula~(\ref{Ml3fin})] 
that analytical continuation of the matrix  $M(z)$ on unphysical 
sheets of energy $z$ is expressed in terms of analytical 
continuation of the truncated scattering matrices $S_l(z)$ or 
$\St_l(z)$ and the half--on--shell Faddeev components $M\abd(z)$ 
taken on the physical sheet.  More precisely, along with  
$S_l(z)$, the final formula~(\ref{Ml3fin}) includes the matrices 
$\bigl( L_0 \bJo M \bigr)(z)$,\,\, $\bigl( L_1\rJ_1 \Psis\Y 
M\bigr)(z)$ and $\bigl( M \bJot L_0  \bigr)(z)$,\,\, $\bigl( 
M\Y\Psi\rJt_1 L_1 \bigr)(z)$.  Here, $l$ is a certain 
multi--index (\ref{lmulti}) and $L=\diag\{ L_0,L_1 \}$ is the 
respective matrix (\ref{Lmatrix}) with  $L_0=l_0$.

In the rest of this section we shall formulate some statements 
(Theorems~\ref{ThLTL}--\ref{ThJ0MYPsiJ1t}) concerning the 
existence of the analytical continuation of the above matrices 
and their domains of holmorphness. In view of shortage of space 
we shall not give here full proofs. Note only that proofs are 
based on analysis~\cite{Faddeev63} of the Faddeev 
equations~(\ref{MFE}).  For all this, one has additionally to 
pay a special attention to studying the domains of holomorphness 
in $z$ of the functions 
\be
\label{Znam}
\left[ p\ad^2+{p\bd'}^2-2c\abd(p\ad,p'\bd)-s\abd^2 z \right]^{-1},
\ee
with one or both arguments $p\ad$ and $p'\bd$ 
situating on the energy shells (\ref{EnergyS3}) or (\ref{EnergyS2}). 
Functions (\ref{Znam}) arise when iterating Eq.~(\ref{MFE}) because 
of the presence of the multiplier 
$\bR_0$ in the operator $-\bt\bR_0\Y$. Also, the functions (\ref{Znam}) 
appear  as a display of singularities (\ref{psiphij}) of the eigenfunctions 
$\psi\ajd,$\,\, $\anum$.

In the case when the arguments 
$p\ad$ and/or $p'\bd$ are taken on the shells 
(\ref{EnergyS2}), \,\, $p\ad=\sqrt{z-\lambda\aid}\,\hat{p}\ad$ and 
$p'\bd=\sqrt{z-\lambda\bjd}\,\hat{p}'\ad$,
the holomorphness domains of the functions (\ref{Znam}) 
with respect to the variable $z$ are described by the following 
plain lemmas. 
\begin{lemma}\label{LEqParab}\hspace*{-0.5em}{\sc .}
For any $\rho\geq 0$,\,\, $-1\leq \eta\leq 1$,  the domain 
\be
\label{DomainParab}
  \Real z > \Frac{\lambda}{c^2}+\Frac{c^2}{4s^2|\lambda|}(\Img z)^2
\ee
contains no roots $z$ of the equation 
\be
\label{quadr0}
 z-\lambda + \rho + 2c\sqrt{z-\lambda}\sqrt{\rho}\:\eta
  -s^2 z = 0,
\ee
with $\lambda < 0,$\,\, $0 < |c| < 1$ and $s^2=1-c^2.$
For any number $z\in\bC$ outside the domain (\ref{DomainParab}) 
one can always find such values of parameters 
$\rho\geq 0$  and  $\eta,$ $-1\leq \eta \leq 1,$
that the 
left--hand part of Eq.~(\ref{quadr0}) becomes equal to zero at the point 
$z$.
\end{lemma}
\begin{lemma}\label{LEqQuadr}\hspace*{-0.5em}{\sc .}
Let the parameters of the equation 
\be
\label{quadr}
 z-\lambda_1 + z - \lambda_2 + 2c\sqrt{z-\lambda_1}\sqrt{z-\lambda_2}\:\eta
  -s^2 z = 0
\ee
be such:\,\, $\eta\in[-1,\, 1]$,\,  
 $\lambda_1 \leq \lambda_2 < 0,$\,\, $0 < c <1 $ and $s^2 =1 -c^2$.
Then the following assertions take place. 

1) If $|\lambda_2|>c^2 |\lambda_1|$
then for all $\eta\in[-1,1]$
Eq.~(\ref{quadr}) has a unique root $z$ and this root is real.
Moreover $z=z_{+}$ if  $\eta\geq 0$,
and $z=z_{-}$ if $\eta\leq 0$ with  
\be
\label{koren}
   z_{\pm}=\Frac{
      (1+c^2-2c^2\eta^2)(\lambda_1+\lambda_2) \pm
     2\sqrt{
           c^2\eta^2[\lambda_1 \lambda_2\, s^4-
        (\lambda_2-\lambda_1)^2 c^2 (1-\eta^2)]     }  }
   {(1+c^2)^2 - 4 c^2 \eta^2}.
\ee
When $\eta$ runs the interval $[-1,1]$, the roots $z_\pm$ fill
the interval $ [z_{\rm lt}, z_{\rm rt}] $ with the ends
\be
\label{zl}
 z_{\rm lt}=\Frac{1}{s^2} [-|\lambda_1|-|\lambda_2|-
 2c\sqrt{|\lambda_1| \cdot |\lambda_2|}]
\ee
and 
\be
\label{zr}
 z_{\rm rt}=\Frac{1}{s^2} [-|\lambda_1|-|\lambda_2|+
 2c\sqrt{|\lambda_1| \cdot |\lambda_2|}], \quad z_{\rm rt}< \lambda_1.
\ee

\noindent 2) If $|\lambda_2|=  c^2 |\lambda_1|$
then Eq.~(\ref{quadr}) has two real roots: 

a) the root $z=\lambda_1$ existing for all $\eta\in [-1,1];$

b) the root $ z=z_{-}$ 
given by~(\ref{koren}) which exists for  $-1\leq \eta\leq 0$ only.

For $-1\leq\eta\leq 1$
these roots together fill the interval $[z_{\rm lt},\lambda_1]$ with 
$ z_{\rm lt}=-|\lambda_1|\left(1+{2c^4}/{s^2}\right). $

\noindent 3) If $|\lambda_2| <  c^2 |\lambda_1|$ then 

a) for $-1\leq\eta\leq\eta^{*},$
$
    \eta^{*}=\Frac{\sqrt{c^2-\rho}\,\sqrt{1-c^2\rho}}{c(1-\rho)},
$
$\rho=\Frac{|\lambda_2|}{|\lambda_1|},$
Eq.~(\ref{quadr}) has two real roots $z_{\pm}$ 
given by (\ref{koren}), which fill the interval 
$[z_{\rm lt}, z_{\rm rt}]$ 
with the ends (\ref{zl}) and (\ref{zr}),  $z_{\rm rt}<\lambda_1;$

b) for $\eta^{*} < \eta \leq 0 $  Eq.~(\ref{quadr})
has two complex roots $z_{\pm}$ described again by Eq.~(\ref{koren}). 
When $\eta$ moves, these roots fill the ellipse centered in the point 
$
   z_c=-|\lambda_1|\left[ 1 +
  \Frac{(c^2-\rho)^2}{s^2(1+c^2)(1+\rho)} \right]. 
$
Half--axes of the ellipse are given by 
$
 a=|\lambda_1| \cdot \Frac{(c^2-\rho)(1-c^2\rho)}
 {  (1+c^2)s^2(1+\rho)  }
$
(along real axis) and 
$
 b=|\lambda_1| \cdot \Frac{(c^2-\rho)(1-c^2\rho)}
 {   (1+c^2)s^2(1-\rho)\sqrt{(1+c^2)^2 - 4c^2\eta^{*2}}  }
$
(along imaginary axis). The right vertex of the ellipse 
is located in the point 
$z^{({\rm e})}_{\rm rt}=z_c+a=$ $-\Frac{|\lambda_1|+|\lambda_2|}{1+c^2}$
situated between  $\lambda_1$ and $\lambda_2$. Its left vertex is 
$z^{({\rm e})}_{\rm lt}=z_c-a < z_{\rm rt}$.
\end{lemma}

Let  $\Pibj$ be the domain in the complex plane $\bC$ with cut 
along the ray $[\lambda_{\rm min},+\infty)$ where the conditions 
(\ref{DomainParab}) with $\lambda=\lambda\bjd$, $c=c\abd$ and 
the inequalities 
\be
\label{Neqsb}
\Real z > \lambda\bjd-s\abd^2 b^2 +\Frac{1}{4 s\abd^2 b^2} (\Img z)^2
\ee
are valid simultaneously for all $\an=1,2,3,$\,\, $\an\neq\bn$.
In the case of the potentials (\ref{vpot}) one has to take 
$b=+\infty$ in (\ref{Neqsb}).

By $\cR_{\an,i;\bn,j}$, $\an\neq\bn$, we denote domain complementary 
in $\bC\setminus[\lambda_{\rm min},+\infty)$ to the set 
filled by the roots of Eq.~(\ref{quadr}) in the case when 
$\lambda_1={\rm min} \{\lambda\aid,\lambda\bjd\}$,
$\lambda_2={\rm max} \{\lambda\aid,\lambda\bjd\}$,
$c=|c\abd|$
and  $\eta=(\hp\ad,\hp'\bd)$
runs the interval $[-1,1]$.
\begin{theorem}\label{ThLTL}  \hspace*{-0.5em}{\sc .}
The matrix integral operator $L'_1\hat{\cT}_{11}(z)L''_1$, \,\,
$z=E\pm i0$,
acting in $\hat{\cH}_1$, allows analytical continuation in $z$
from rims of the ray $E\in(\lambda,+\infty)$,\,\,
$
\lambda=\!\! \Max{\mbox{   \scriptsize
                  $\begin{array}{c}
            l'_{\gn,k}\neq 0, \\
            l''_{\gn,k}\neq 0
                        \end{array}$
                   }
	     }
\!\! \lambda_{\gn,k}
$,
on the domain 
\be
\label{Pil1hol}
\Pi^{({\rm hol})}_{l'l''}=
\bigl[ \Bigcap_{               \mbox{\scriptsize
                        $\begin{array}{c}
                      l'\aid\neq 0   \\
                      l''\bjd\neq 0
                       \end{array}$  }
                }
\cR_{\an,i;\,\, \bn,j}                              \bigr]
\Bigcap
\bigl[ \Bigcap_{               \mbox{\scriptsize
                        $\begin{array}{c}
                               l'_{\gn,k}\neq 0,   \\
                               l''_{\gn,k}\neq 0
                       \end{array}$  }
                }
\Pi_b^{(\gn,k)}  \bigr]
\setminus\overline{\sigma(H)}
\ee
where 
 $l'_1=\diag\left( l'_0,l'_{1,1},...,l'_{1,n_1},\right.$
 $l'_{2,1},...,l'_{2,n_2},$
 $\left. l'_{3,1},...,l'_{3,n_3}\right),$
 $l''_1=\diag\left( l''_0,l''_{1,1},...,l''_{1,n_1},\right.$
 $l''_{2,1},...,l''_{2,n_2},$
 $\left. l''_{3,1},...,l''_{3,n_3}\right),$
with  $l'_0=l''_0=0.$
The nontrivial kernels 
$
\left( L'_1\hat{\cT}_{11}(z)L''_1  \right)_{\an,i;\,\, \bn,j}
(\hp\ad,\hp\bd',z),
$
\,\, $ l'\aid\neq 0$,\,   $ l''\bjd\neq 0, $
turn into functions holomorphic concerning  
$z\in\Pi^{({\rm hol})}_{l'l''}$
and real--analytic with respect to  $\hp\ad,\hp\bd'\in S^2$.
\end{theorem}
\begin{note}\label{NotePiSym}  \hspace*{-0.5em}{\sc .}
{\rm  The domains 
$\Pi^{({\rm hol})}_{l'l''}$ and $\Pi^{({\rm hol})}_{l''l'}$ coincide,
$\Pi^{({\rm hol})}_{l'l''}=\Pi^{({\rm hol})}_{l''l'}$. }
\end{note}

If $l'=l''=l$, we use for $\Pi^{({\rm hol})}_{l'l''}$ 
the notation $\Pi^{({\rm hol})}_{l}$,\,\,
\be
\label{Pilh1}
\Pi^{({\rm hol})}_{l}=\Pi^{({\rm hol})}_{ll}.
\ee
\begin{theorem}\label{ThMYPsiJt}  \hspace*{-0.5em}{\sc .}
Let $L_0=l_0=0.$ Then the matrices 
 $\bigl(  M\Y\Psi\rJt_1 L_1\bigr)(z)$
and 
 $\bigl( L_1 \rJ_1 \Psis \Y M \bigr)(z)$,
 $z=E\pm i0$,
allow analytical continuation in $z$ from rims of the ray 
$E\in(\lambda,+\infty)$,\,\,
$\lambda=\Max{(\bn,j):\, l\bjd\neq 0} \lambda\bjd$,
on the domain 
$\Pilh\setminus\overline{\sigma(H)}$
as a bounded for 
$ z\not\in[\lambda_{\rm min}, +\infty)$
operator--valued functions of variable $z$,\,\,
$\bigl(  M\Y\Psi\rJt_1 L_1\bigr)(z)$:
$\hat{\cH}_1\rightarrow\cG_0$  and 
 $\bigl( L_1 \rJ_1 \Psis \Y M \bigr)(z)$:
 $\cG_0\rightarrow \hat{\cH}_1$.
\end{theorem}

Continuing the half--on--shell matrices 
$(\bJo M)(z),$\,\,\, $(M\bJot)(z),$\,\, $z=E\pm i0,$\,\,\, $E>0,$
into domain of complex $z$ is considered in a sense of 
distributions over $\Osix$. For example of 
$M\bJot$ we consider continuation of the bilinear form 
$$
{\bigl(F,\, (M\bJot)(E\pm i0) \bigr)\equiv}\\
\Sum_{\an,\bn}\Int_\Rs dP \Int_{S^5}d\hP'
\, F\ad(P)M\abd(P,\pm\sqrt{E}\,\hP',E\pm i0)\, f\bd(\hP')
$$
where $F=(F_1,F_2,F_3)$ with $F\ad\in\Osix$ and  $f=(f_1,f_2,f_3)$ with 
$f\ad\in\hat{\cH}_0$.

When constructing continuation of this form and that for
$(\bJo M)(E\pm i0)$ we base on two simple statements 
concerning the domains of holomorphness of the function (\ref{Znam}) 
in the case when argument $P'$ belongs to the three--body 
energy shell (\ref{EnergyS3}) and therefore 
$p'\bd=\sqrt{z}\nu'\hp'\bd$ with 
$\nu'\in[0,\, 1]$.
\begin{lemma}\label{LEqQuadrPrime}\hspace*{-0.5em}{\sc .}
Let in the equation 
$
\rho + z\nu' + 2c\sqrt{z}\sqrt{\nu'}\sqrt{\rho}\eta -s^2 z = 0
$,
the parameters  $\nu'$ and $\eta$ run the intervals  
$0\leq\nu'\leq 1$ and  $ -1 \leq \eta \leq 1 $ respectively,
and $c > 0,$  $ s^2 =1-c^2$,  $ z\in\bC$ be fixed. Then 
the roots $\rho$ of the above equation 
fill the set consisting of the line segment $[0,z]$ on 
the complex plane $\bC$ and the circle centered in the origin, 
the radius of which being equal to $c^2 |z|.$
\end{lemma}
\begin{lemma}\label{LEq23}\hspace*{-0.5em}{\sc .}
Let the parameters of the equation 
\be
\label{QEq23}
z-\lambda + z\nu +
2c\sqrt{z}\sqrt{z-\lambda}\sqrt{\nu}\eta -s^2 z = 0,
\ee
satisfy the conditions $\nu\in[0,1]$, $\eta\in[-1,1]$, $\lambda<0$,
$c\in(0,1)$ and $ s^2 =1-c^2$. Then if 
$\nu$ and $\eta$ run the above ranges, the roots $z$ 
of Eq.~(\ref{QEq23}) fill the ray 
$\left(-\infty, {\lambda}/({1+c^4})\right]$ and the circle centered 
in the point 
$z_c={\lambda}/({1-c^4})$,
radius of which equals to ${c^2\lambda}/({1-c^4})$.
\end{lemma}

Let $\tilde{\Pi}_b^{(0)\pm}$,\,\,
$\tilde{\Pi}_b^{(0)\pm}\!\subset\!\bC^\pm $, be the domains 
complementary in $\bC^\pm$ to the totality of 
circles having radii $r={c\abd^2|\lambda\ajd|}/({1-c\abd^4})$ and 
centered in the points 
$z_c={\lambda\ajd}/({1-c\abd^4})$
where $\an,\bn=1,2,3,$\, $\bn\neq\an$, and $j=1,2,...,n\ad$.
In the case of the potentials (\ref{vpotb})
the domains $\tilde{\Pi}_b^{(0)\pm}$ must satisfy extra conditions 
\be
\label{CondbRes0}
\Real z > -\Frac{|s\abd|^2 b^2}{(1+|c\abd|)^2}
+\Frac{(1+|c\abd|)^2}{4|s\abd|^2 b^2}\, (\Img z)^2.
\ee
for all $\an,\bn=1,2,3,$\, $\bn\neq\an$.

Utilizing Lemmas~\ref{LEqQuadrPrime} and~\ref{LEq23}
one can prove the following 
\begin{theorem}\label{ThMJ0tJ0M}  \hspace*{-0.5em}{\sc .}
Kernels of the matrices  $(M\bJot)(z)$
and $(\bJo M)(z)$, \,\, $z=E\pm i0$,\,\, $E>0$,
allow analytical continuation in $z$
on the domains, respectively 
$\tilde{\Pi}_b^{(0)+}$ and $\tilde{\Pi}_b^{(0)-}$, \,
$\tilde{\Pi}_b^{(0)\pm}\subset\bC^\pm$.
The continuation of kernels of the matrices 
$\left( \cQ^{(n)}\bJot  \right)(z)$,
and  $\left( \bJo\cQ^{(n)}\right)(z)$,\,\, $n\leq 3$,
included in the representation (\ref{MQW}) for  $M(z)$ 
has to be understood in a sense of distributions over  $\Osix$. 
At the same time the kernels 
\be
\label{WJ0t}
\begin{array}{c}
\cF\abd(P,\sqrt{z}\hP',z),\,\,
\cI_{\an,j;\bn}(p\ad,\sqrt{z}\hP',z),\,\,  \\
\cJ_{\an;\bn,k}(P,\sqrt{z}\sqrt{\nu'}\,\hp'\bd,z)\,\,
\mbox{ and }\, \cK_{\an,j;\bn,k}(p\ad,\sqrt{z}\sqrt{\nu'}\,\hp'\bd,z)
\end{array}
\ee
$$
\an,\bn=1,2,3,\,\, j=1,2,...,n\ad,\,\, k=1,2,...,n\bd,\,\,
$$
of the matrices $\bigl(\cQ\un\bJot\bigr)(z)$,\,\, $n\geq 4$,\,\, 
and $(\cW\bJot)(z)$ as well as the kernels 
\be
\label{J0W}
\begin{array}{c}
\cF\abd(\sqrt{z}\hP,P',z),\,\,
\cI_{\an,j;\bn}(\sqrt{z}\sqrt{\nu}\,\hp\ad,P',z),\,\,  \\
\cJ_{\an;\bn,k}(\sqrt{z}\hP,p'\bd,z)\,\,
\mbox{ and }\, \cK_{\an,j;\bn,k}(\sqrt{z}\sqrt{\nu}\,\hp\ad,p'\bd,z)
\end{array}
\ee
of the matrices 
$\bigl(\bJo\cQ\un\bigr)(z)$,\,\, $n\geq 4$,\,\, and  $(\bJo\cW)(z)$
can be continued on the domains $\tilde{\Pi}_b^{(0)\pm}$
as usual holomorphic functions of variable $z$.
Being H\"older functions of variables 
$\hP'\in S^5$ or $\sqrt{\nu'}\hp'\bd$,\,\, $0\leq\nu'\leq 1$,\,\,
$\hp'\bd\in S^2$\,\,
($\hP\in S^5$ or $\sqrt{\nu}\hp\ad,$
\,\, $0\leq\nu\leq 1$,\,\,
$\hp\ad\in S^2$) with index $\mu'\in(0,\,\,1/8)$,
the kernels  (\ref{WJ0t})\,\,
(kernels (\ref{J0W}))
considered as functions of $P\in\Rs,$ $p\ad\in\Rt$
($P'\in\Rs$, $p'\bd\in\Rt$), can be embedded in their totality in 
$\cB(\theta,\mu)$\,\, with $\theta$ and $\mu$, the arbitrary 
numbers such that $\theta\in (3/2,\,\theta_0)$
\,\, and $\mu\in(0,\, 1/8)$.
At  $|\Img z|\geq \delta > 0$
one can take $\mu=1$.
\end{theorem}

Let us comment the assertion of the theorem for example of the matrices 
$\bigl(M\bJot \bigr)(z)$. Note in particular that 
continuation on $\tilde{\Pi}_b^{(0)\pm}$  of the form 
$\left(F,\bigl(\cQ^{(0)}\bJot \bigr)(z)f\right)=$
$\sum\ad\left(F\ad,\bigl(\bt\ad\bJot \bigr)(z)f\ad\right)$
is described by the equalities 
\be
\label{FtJ0f}
\begin{array}{c}
\bigl(F\ad,(\bt\ad\rJot)(z)f\ad \bigr)
\equiv \Int_\Rt dk\ad \Int_{S^2}d\hk'\ad      \Int_{S^2}d\hp'\ad
\Int_0^{\pi/2}   d\omega'\ad\sin^2\omega'\ad
\cos^2\omega'\ad  \times   \\
\times\, t\ad(k\ad,\sqrt{z}\cos\omega'\ad\hk'\ad, z\cos^2\omega'\ad )
 F\ad(k\ad, \pm\sqrt{z}\sin\omega'\ad\hp'\ad)\cdot
f\ad(\omega'\ad,\hk'\ad,\hp'\ad),           \phantom{\Int}
\end{array}
\ee
where $\omega'\ad,\hk'\ad,\hp'\ad$ are the hyperspherical 
coordinates~\cite{MF} of the point 
$\hP'\in S^5$, $\omega'\ad\in\left[0,\pi/2\right]$,
$\hk'\ad,\hp'\ad \in S^2$.
Note also that 
$\hP'=\lbrace \cos\omega'\ad\hk'\ad,\,\,\sin\omega'\ad\hp'\ad\rbrace$
and 
$
d\hP'=\sin^2\omega'\ad\cos^2\omega'\ad d\omega'\ad d\hk'\ad d\hp'\ad
$
is a measure on $S^5$.

The analytical continuation on $\tilde{\Pi}_b^{(0)\pm}$ of the form 
$\left(F,\bigl(\cQ^{(1)}\bJot \bigr)(E\pm i0)f\right)$ 
is given by 
\be
\label{Q1Distr12}
\left(F,\bigl(\cQ^{(1)}\bJot \bigr)(z)f\right)=
\Sum_{\an,\bn,\,\, \an\neq\bn }
Q_{1,\an\bn}^\pm(z)+Q_{2,\an\bn}^\pm(z)
\ee
where 
\be
\label{Q1cont}
\begin{array}{c}
Q_{1,\an\bn}^\pm(z)=\pm\Frac{\sqrt{z}}{4}\Frac{1}{|s\abd|}
\Int_{\Rt} d{k}\ad \Int_{S^2} d\hat{p}\ad
  \Int_{S^2} dk'\bd \Int_{S^2} d\hat{p}'\bd
  \Int_0^{1} d\nu\, \sqrt{\nu} \cdot
  \Int_0^{1} d\nu'\, \sqrt{\nu'}\sqrt{1-\nu'} \times   \\
%
                  \phantom{\cdot}     \\
  \times\, \Frac{ F\ad({k}\ad,\sqrt{z}\sqrt{\nu}\hat{p}\ad) \cdot
 f\bd(\sqrt{1-\nu'}\hat{k}'\bd,\sqrt{\nu'}\hat{p}'\bd) }
   {\nu+\nu'-2c\abd\sqrt{\nu}\sqrt{\nu'}\,(\hat{p}\ad,\hat{p}'\bd)
   -s\abd^2 \mp  i0}              \times  \\
  \times\, t\ad({k}\ad,
   \tk\bu\ad(\sqrt{z}\sqrt{\nu}\hat{p}\ad,\sqrt{z}\sqrt{\nu'}\hat{p}'\bd),
   z(1-\nu))\, \times       \phantom{\Int}         \\
  \times\,
t\bd(\tk\au\bd(\sqrt{z}\sqrt{\nu'}\hat{p}'\bd,\sqrt{z}\sqrt{\nu}\hat{p}\ad),
\sqrt{z}\sqrt{1-\nu'}\hat{k}'\bd, z(1-\nu'))
\end{array}
\ee
and 
\be
\label{Q2cont}
\begin{array}{c}
Q_{2,an\bn}^\pm(z)=\pm\Frac{1}{4}\cdot\Frac{1}{|s\abd|}
\Int_{\Rt} d{k}\ad \Int_{S^2} d\hat{p}\ad
  \Int_{S^2} dk'\bd \Int_{S^2} d\hat{p}'\bd
  \Int_{\Gamma_z^\pm} d\rho\, \sqrt{\rho} \cdot
  \Int_0^{1} d\nu'\, \sqrt{\nu'}\sqrt{1-\nu'} \times   \\
                  \phantom{\cdot}     \\
  \times\, \Frac{ F\ad({k}\ad,\pm\sqrt{\rho}\hat{p}\ad) \cdot
 f\bd(\sqrt{1-\nu'}\hat{k}'\bd,\sqrt{\nu'}\hat{p}'\bd) }
{\rho+z\nu'-2c\abd\sqrt{z}\sqrt{\rho}\sqrt{\nu'}\,(\hat{p}\ad,\hat{p}'\bd)
   -s\abd^2 z}  \,            \times  \\
  \times\, t\ad({k}\ad,
   \tk\bu\ad(\pm\sqrt{\rho}\hat{p}\ad,\sqrt{z}\sqrt{\nu'}\hat{p}'\bd),
   z-\rho)\, \times       \phantom{\Int}         \\
  \times\,
t\bd(\tk\au\bd(\sqrt{z}\sqrt{\nu'}\hat{p}'\bd,\pm\sqrt{\rho}\hat{p}\ad),
\sqrt{z}\sqrt{1-\nu'}\hat{k}'\bd, z(1-\nu')).
\end{array}
\ee
Here, by $\Gamma_z^+$\,\, ($\Gamma_z^-$) we understand a path of integration 
beginning at $z$ and going clockwise (counterclockwise) along 
the circumference $C_{|z|}$ having radius $|z|$ and centered in the origin. 
After the path crosses the real axis, it goes further along this one  
so that the rest of $\Gamma_z^+$\,\, ($\Gamma_z^-$) consists of the points   
$\rho=\lambda + i0$\, ($\rho=\lambda + i0$),\, $\lambda\in(|z|,+\infty).$

Boundaries of the holomorphness domains $\tilde{\Pi}_b^{(0)\pm}$ 
of the form $\left(F,\bigl(\cQ^{(1)}\bJot \bigr)(z)f\right)$
are found as a matter of the fact, 
from those requirements that the poles of 
$T$--matrices $t\ad(\,\cdot\, ,\,\cdot\, ,z(1-\nu))$
and  $t\bd(\,\cdot\, ,\,\cdot\, ,z(1-\nu'))$  
which are present in the integral (\ref{Q1cont}), have not to manifest 
itself in above domains. Also, we require the same from the poles of 
$T$--matrices $t\ad(\,\cdot\, ,\,\cdot\, ,z-\rho)$
which are present in the integral (\ref{Q2cont}).
If $z\not\in(-\infty,\lambda_{\rm max}]$
then the appearance conditions 
$
z(1-\nu)=\lambda\ajd,\,\,\, j=1,2,...,n\ad,\quad
z(1-\nu')=\lambda_{\bn,k},\,\,\, k=1,2,...,n\bd,
$
for the poles of the $T$--matrices
$t\ad(\,\cdot\, ,\,\cdot\, ,z(1-\nu))$
and $t\bd(\,\cdot\, ,\,\cdot\, ,z(1-\nu'))$,
are valid for no $\nu,\nu'\in[0,\,1]$.
The appearance conditions $z-\rho=\lambda\ajd,$\,\, $j=1,2,...,n\ad,$ 
of the  poles of $t\ad(\,\cdot\, ,\,\cdot\, ,z-\rho)$
may be realized if only the contours $\Gamma_z^\pm$
include into itself more than one fourth of the circumference $C_{|z|}$.
However their contribution to $Q_{2,\abd}^\pm(z)$
arising when the points $\rho=z-\lambda\ajd$ cross contours 
$\Gamma_z^\pm$, may be always taken into account 
using the residue theorem.
We shall not present here respective formulae. Note only that 
taking of residues in the points $\rho=z-\lambda\ajd$
transforms the minor three--body pole singularities
of the integrand of $Q_{2,\an\bn}^\pm(z)$ into those of the type 
$
%
(z-\lambda\ajd+z\nu'-2c\abd\sqrt{z}\sqrt{z-\lambda\ajd}
\sqrt{\nu'}\eta-s\abd^2 z)^{-1}.
$
Location of such singularities is described 
by Lemma~\ref{LEq23}.

The iteration $\cQ^{(2)}(z)$ kernels 
$\cF\abd(P,P',z)$,\,\, 
$\cI_{\an,j;\bn}(p\ad,P',z),$\,\,
$\cJ_{\an;\bn,k}(P,p'\bd,z)$,\,\,
and 
\newline
$\cK_{\an,j;\bn,k}(p\ad,p'\bd,z)$,\, $P,P'\in\Rs$,\,\, 
$p\ad,p'\bd\in\Rt$, have more weak 
singularities~\cite{Faddeev63}, \cite{MF} than the 
$\cQ^{(1)}(z)$ components. When continuing the form 
$\left(F,\bigl(\cQ^{(2)}\bJot \bigr)(z)f\right)$ we get for it 
the representations which differ from 
(\ref{Q1Distr12})--(\ref{Q2cont}) mainly in replacement of the 
distributions 
$
   \{ z(\nu+\nu'-
   2c\abd\sqrt{\nu}\sqrt{\nu'}(\hp\ad,\hp'\bd) -s\abd^2 \mp i0)\}^{-1}
$,\,\,
$0\leq \nu \leq 1,$ \,\, $0\leq \nu'\leq 1,$
with functions singular as 
\be
\label{It20Sing}
\Frac{1}{z\, |c_{\an\gn}\nu\hp\ad-c_{\bn\gn}\nu'\hp'\bd|}\cdot
\mbox{\large $\ln$}
\Frac{   \sqrt{s_{\an\gn}^2(1-\nu^2)}+\sqrt{s_{\bn\gn}^2(1-{\nu'}^2)}+
|c_{\an\gn}\nu\hp\ad-c_{\bn\gn}\nu'\hp'\bd|   }
{   \sqrt{s_{\an\gn}^2(1-\nu^2)}+\sqrt{s_{\bn\gn}^2(1-{\nu'}^2)}-
|c_{\an\gn}\nu\hp\ad-c_{\bn\gn}\nu'\hp'\bd|   }
\ee
The kernels $\cF\abd(P,P',z)$
$\cI_{\an,j;\bn}(p\ad,P',z),$\,\,
$\cJ_{\an;\bn,k}(P,p'\bd,z)$,\,\,
and $\cK_{\an,j;\bn,k}(p\ad,p'\bd,z)$\,
of the iteration $\cQ^{(3)}(z)=\left( -\bt(z)\bRo(z)\Y\right)^3 \bt(z)$
are still singular. Though their singularities are weak, 
continuation of the kernels $\left( \cQ^{(3)}\bJot  \right)(z)$
on the domains $\tilde{\Pi}_b^{(0)\pm}$
we understand as before in a sense of distributions over $\Osix$. 
So, we realize it following the same scheme 
as for the continuation of 
$\left( \cQ^{(1)}\bJot  \right)(z)$
and $\left( \cQ^{(2)}\bJot  \right)(z)$.
\begin{theorem}\label{ThJ0TJ0t}  \hspace*{-0.5em}{\sc .}
The matrix $\bigl(\bJo M\bJot \bigr)(z)$
(the operator  $\bigl(\rJo T\rJot \bigr)(z)$)
admits the analytical continuation in  $z$
from the rims of the cut $z=E\pm i0$,\,\, $E> 0$,
on the domains $\tilde{\Pi}_b^{(0)\pm}\in\bC^\pm$
as a bounded operator in $\hat{\cG}_0$
(in $\hat{\cH}_0$).
For all this $\bigl(\bJo M\bJot \bigr)(z)$,\,\, $z\in\tilde{\Pi}_b^{(0)\pm}$,
admits the representation {\em [cf.~(\ref{MQW})]}\,\,
$
\bigl(\bJo M\bJot \bigr)(z)=\Sum_{n=0}^3 \bigl(\bJo \cQ\un\bJot \bigr)(z)
+\bigl(\bJo \cW \bJot \bigr)(z).
$
\, The operators $\bigl(\bJo \cQ^{(0)}\bJot \bigr)(z)$
and $\bigl(\bJo \cQ^{(1)}\bJot \bigr)(z)$
are bounded matrix operators in $\hat{\cG}_0$
with singular kernels.
Having weakly singular kernels the matrices 
$\bigl(\bJo \cQ\un\bJot \bigr)(z)$,\,\,\, $n=2,3$,
are compact operators in $\hat{\cG}_0$.
To that end kernels of matrix $\bigl(\bJo \cW \bJot \bigr)(z)$
are H\"older functions of their arguments with the index 
$\mu\in\bigl(0,\,\,1/8 \bigr)$.
\end{theorem}

As a comment to this theorem we present explicit formulae 
for the kernels of the operators $\bigl(\bJo \cQ^{(0)}\bJot \bigr)(z)$
and $\bigl(\bJo \cQ^{(1)}\bJot \bigr)(z)$.

The first of them have the form 
$\bigl(\bJo \cQ^{(0)}\bJot \bigr)\abd(\hP,\hP',z)=$
$\delta\abd\bigl(\rJo\bt\ad\rJot \bigr)(\hP,\hP',z)$,
$\an,\bn=1,2,3,$ where 
\be
\label{JTJS}
\begin{array}{c}
{(\rJo\bt\ad\rJot)(\hP,\hP',z)=
t\ad(\sqrt{z}\cos\omega\ad\hk\ad,\sqrt{z}\cos\omega{'}\ad\hk{'}\ad,
z\cos^2\omega\ad)\,\times} \\
\phantom{\cdot}\\             \times\,
\delta(\sqrt{z}\sin\omega\ad\hp\ad - \sqrt{z}\sin\omega{'}\ad\hp'\ad).
\end{array}
\ee
Here,  $\omega\ad,\hk\ad,\hp\ad$ and $\omega'\ad,\hk'\ad,\hp'\ad$
are coordinates of the points 
$\hP=\lbrace k\ad,p\ad\rbrace$ and $\hP'=\lbrace k'\ad,p'\ad\rbrace$
on hypersphere $S^5$.
We mean here that 
\be
\label{deltps}
\delta(\sqrt{z}\sin\omega\hp - \sqrt{z}\sin\omega{'}\hp')=
{\rm Sign}\,\Img z\cdot
\Frac{\delta(\hp,\hp')\delta(\omega-\omega')}
{(\sqrt{z})^3 \sin^2\omega\cos\omega},
\ee
where $\delta(\hp,\hp')$  is the kernel of the 
identity operator in $L_2(S^2)$.
The denominator  $(\sqrt{z})^3 \sin^2\omega\cos\omega$
of the right--hand part of Eq.~(\ref{deltps})
represents analytical continuation of the Jacobian for 
respective replacement of variables.

Therefore  the operator 
$\bigl(\rJo\bt\ad\rJot\bigr)(z)$ 
acts at $\Img z\neq 0$ on $f\in\hat{\cH}_0$ as 
\begin{eqnarray}
\nonumber
\lefteqn{ \left(\bigl( \rJo\bt\ad\rJot \bigr)(z)f \right)(\hP)=
\Frac{{\rm Sign}\,\Img z}{\bigl(\sqrt{z} \bigr)^3}\cdot
\Int_{S^2} d\hk'\ad\,\, \times}  \\
\label{J0t2J0t}
&& \times\,\,
t\ad(\sqrt{z}\cos\omega\ad\hk\ad,\,\sqrt{z}\cos\omega\ad\hk'\ad,\,
z\cos^2\omega\ad)
f(\cos\omega\ad\hk'\ad,\sin\omega\ad\hp\ad).
\end{eqnarray}

The operators $\bigl(\bJo\cQ^{(1)}\bJot \bigr)(z)$, \,\,
$z\in\tilde{\Pi}_b^{(0)\pm}$,
have the kernels 
$$
%
\bigl(\bJo\cQ^{(1)}\bJot \bigr)\abd(\hP,\hP',z)=
\Frac{1}{z}\cdot\Frac{1-\delta\abd}{|s\abd|}\cdot
 \Frac{ t\ad\bigl(k\ad ,k\ad\bu ,z(1-\nu) \bigr)
\,\,\, t\bd\bigl(k\bd\au ,k'\bd, z(1-\nu') \bigr)  }
{\nu+\nu'-2c\abd\sqrt{\nu}\sqrt{\nu'}(\hp\ad,\hp'\bd)-s\abd^2\mp i0},
%
$$
where $k\ad=\sqrt{z}\sqrt{1-\nu}\hk\ad,$\,\,
$k'\bd=\sqrt{z}\sqrt{1-\nu'}\hk'\bd,$\,\,
$
k\ad\bu=\tk\ad\bu\bigl(\sqrt{z}\sqrt{\nu}\hp\ad,\,
\sqrt{z}\sqrt{\nu'}\hp'\bd \bigr)
$
and 
$
k\bd\au=\tk\bd\au\bigl(\sqrt{z}\sqrt{\nu'}\hp'\bd,\,
\sqrt{z}\sqrt{\nu}\hp\ad \bigr). 
$
At the same time $\nu=\sin^2\omega\ad$ and $\nu'=\sin^2\omega'\bd$.

Main singularities of the kernels 
$\bigl(\bJo\cQ^{(2)}\bJot \bigr)\abd(\hP, \hP',z)$ in $\hP$, $\hP'$ 
are described by Eqs.~(\ref{It20Sing}). Singularities of the kernels 
$\bigl(\bJo\cQ^{(3)}\bJot \bigr)\abd(\hP, \hP',z)$ are more weak. 
\begin{theorem}
\label{ThJ0MYPsiJ1t}  \hspace*{-0.5em}{\sc .}
The operators $\bigl(\bJo M\Y\Psi\rJt_1 \bigr)(z):$
$\hat{\cH}_1\rightarrow\hat{\cG}_0$,
\,\,\,
$\bigl(\rJ_1\Psis\Y M\bJot \bigr)(z):$
$\hat{\cG}_0\rightarrow\hat{\cH}_1$,\,\,\,
$\hat{\cT}_{01}(z):$ $\hat{\cH}_1\rightarrow\hat{\cH}_0$, \,
and 
$\hat{\cT}_{10}(z):$ $\hat{\cH}_0\rightarrow\hat{\cH}_1$ \,
admit the analytical continuation from rims of the cut 
$z=E\pm i0,$\,\, $E>0$,
onto the domains $\Pi_b^{(0)\pm}\subset\bC^\pm$
including the points 
$z\in\tilde{\Pi}_b^{(0)\pm}\Bigcap_{\bn,j}\Pi_b^{(\bn,j)}$
satisfying the additional conditions 
$$
\Real z> \Frac{|s_{\bn\gn}|^2}{(1+|c_{\bn\gn}|)^2}\,\lambda\bjd
+\Frac{(1+|c_{\bn\gn}|)^2}{4|s_{\bn\gn}|^2 |\lambda\bjd|}(\Img z)^2.
$$
for any $\bn,\gn=1,2,3,$\,\, $\bn\neq\gn,$ and  $j=1,2,...,n\bd.$
For all $z\in\Pi_b^{(0)\pm}$ 
including the boundary points $z=E\pm i0$,\,\, $E>0$,
these operators are compact.
\end{theorem}

Later, we shall use the notation 
\be
\label{Pil0hol}
\Pi_{l^\pm}^{({\rm hol})}\equiv\Pi_b^{(0)\pm}\bigcap
\Pi_{l^{(1)}}^{({\rm hol})},
\ee
where $l^\pm=(l_0^\pm,l_{1,1},...,l_{1,n_1},l_{2,1},...,
l_{2,n_2},l_{3,1},...,l_{3,n_3})$
with  $l_0^\pm=\pm 1$, 
 $l\ajd=1,$\,\,$\anum$, 
and  $l^{(1)}=$ $(0,\,\,l_{1,1},...,l_{1,n_1},$ $l_{2,1},...,$
$l_{2,n_2},$ $l_{3,1},...,$ $l_{3,n_3})$ with the same $l\ajd$ as 
$l^\pm$.
Remember that the sets 
$\Pi_{l^{(1)}}^{({\rm hol})}\equiv\Pi_{l^{(1)} l^{(1)}}^{({\rm hol})}$
were defined by Eqs.~(\ref{Pil1hol}).

As follows from Theorems~\ref{ThLTL}, \ref{ThJ0TJ0t} 
and~\ref{ThJ0MYPsiJ1t}, the total three--body scattering matrix 
$S(z)$,\,\, $z=E\pm i0$,\,\, $E>0$, admits the analytical 
continuation as a holomorphic operator--valued function, $S(z):$ 
$\hat{\cH}_0\oplus\hat{\cH}_1\rightarrow\hat{\cH}_0\oplus\hat{\cH}_1$,
on the domain  $\Pi_{l^+}^{({\rm hol})}\subset\bC^+.$ For any 
$z\in\Pi_{l^+}^{({\rm hol})}$ the operator $S(z)$ is bounded. In 
equal degree the same is true for $\St(z)$.

\section{\hspace*{-1em}. 
                         DESCRIPTION OF (PART OF) THE THREE--BODY 
\newline 
                         RIEMANN SURFACE }
\label{SRiemannSurface}
By the {\em three--body energy Riemann surface} we mean 
the Riemann surface of the kernel $R(P,P',z)$ of the Hamiltonian $H$ 
resolvent $R(z)$ considered as a function of parameter $z$, the energy 
of three--body system. 

One has to expect this surface as well as that of the free Green 
function $R_0(P,P',z)$ to consist of infinite number of sheets 
already because the threshold $z=0$ is a logarithmic branching 
point.  Actually the Riemann surface of $R(P,P',z)$ is much more 
complicated than that of  $R_0(P,P',z)$ because besides  $z=0$ 
it has a lot of additional branching points. For example the 
pair thresholds $z=\lambda\ajd$, $\anum$, become square root 
branching points of this surface.  Also, the resonances of the 
pair subsystems turn into such points.  Extra branching points 
are generated by the boundaries of supports of the function 
(\ref{Znam}) singularities which were described in 
Lemmas~\ref{LEqParab}, \ref{LEqQuadr} and \ref{LEq23}.

In the present paper we restrict ourselves 
to consideration of a ``small'' part of the total three--body Riemann 
surface for which we succeeded to find the explicit representations 
expressing analytical continuation of the Green function 
$R(P,P',z)$, \,\, the kernels of the matrix $M(z)$, 
as well as the scattering matrix $S(z)$, in terms of the physical sheet 
[see the formulae respectively, (\ref{Ml3fin}), (\ref{Slfin}) and
(\ref{R3l})].  Namely, in the Riemann surface of $R(P,P',z)$
we consider two neighboring {\em ``three--body''} unphysical sheets 
immediately joint with the physical one along the {\em
three--body} branch of continuous spectrum 
$[0,\,\,+\infty)$.  Besides, we examine all the {\em
``two--body''} unphysical sheets, i.e. the sheets where parameter 
$z$ may be carried if the rounds of two--body thresholds 
$z=\lambda\ajd$, $\anum$, are permitted but 
the crossing of the ray $[0,\,\,+\infty)$ is forbidden.  
Evidently, the part of the three--body surface described includes all 
the sheets neighboring with physical one.  
The above sheets are of most interest in applications.  

A concrete description of the part under consideration we give 
using the auxiliary vector--function 
$\sff(z)=(\sff_0(z),$ $\sff_1(z),$ $\sff_2(z),$ $\sff_3(z))$,
where 
$\sff_0(z)=\ln z$ 
and
$\sff\ad(z)$, $\an=1,2,3$,
are again vector--functions,  
$
\sff\ad(z)=((z-\lambda_{\an,1})^{1/2},
(z-\lambda_{\an,2})^{1/2},...,(z-\lambda_{\an,n\ad})^{1/2}).
$

The Riemann surface of $\sff(z)$ consists of infinite number of 
the copies of the complex plane $\bC'$ cut along the ray 
$[\lambda_{\rm min},+\infty)$. These sheets are sticked together 
in a suitable way along rims of the cut segments between 
neighboring points in the set of thresholds $\lambda\ajd,$ 
$\anum$, and $\lambda_0=0.$ The sheets $\Pi_{l_0 l_1 l_2 l_3}$ 
of this surface are identified by the indices of branches of the 
functions $\sff_0(z)=\ln z$ and 
$\sff\ajd(z)=(z-\lambda\ajd)^{1/2}$ in such a manner that  $l_0$ 
is integer and $l\ad$, $\an=1,2,3$ are multi--indices, 
$l\ad=(l_{\an,1},l_{\an,2},...,l_{\an,n\ad})$, $l\ajd=0,1$.  For 
the main branch of the function $\sff\ajd(z)$, $\anum$, we take 
$l\ajd=0$, and otherwise $l\ajd=1$. In the case if there exist 
coinciding thresholds i.e.  $\lambda_{\an,i}=\lambda_{\bn,j}$ at 
$\an\ne \bn$ and/or $i\ne j$ (this means that discrete spectra 
of the pair Hamiltonian coincide partly though for two pair 
subsystems or though one of the pair subsystems has a multiple 
discrete spectrum) then on the each sheet  $\Pi_{l_0 l_1 l_2 
l_3}$ indices $l_{\an,i}$ and  $l_{\bn,j}$ coincide, too,  
$l_{\an,i}$=$l_{\bn,j}$.  As $l_0$ we choose the number of the 
function $\ln z$ branch, $\ln z=\ln|z|+i\varphi_0+i 2\pi l_0$ 
with $\varphi_0$, the argument of $z$, $z=|z|{\rm 
e}^{i\varphi_0},$ $\varphi_0\in[0,2\pi)$.  Sheets $\Pi_{l_0 l_1 
l_2 l_3}$ are sticked together (along rims of the cut) in such a 
way that if parameter $z$ going from the sheet $\Pi_{l_0 l_1 l_2 
l_3}$ crosses segment of line between two neighboring thresholds 
$\lambda_{\an,i}$ and $\lambda_{\bn,j}$, 
$\lambda_{\an,i} < \lambda_{\bn,j}$ (or $\lambda_{\rm max}$ and 
$\lambda_0$) than it comes to the sheet $\Pi_{l'_0 l'_1 l'_2 l'_3}$,
with indices  
$l_{\gamma,k}$ corresponding to $\lambda_{\gamma,k}\leq\lambda_{\alpha,i}$ 
$(\lambda_{\gamma,k}\leq\lambda_{\rm max})$ which change by 1.  
For all this if $l_{\gamma,k}=0$ then $l'_{\gamma,k}=1$; if 
$l_{\gamma,k}=1$ then $l'_{\gamma,k}=0$.  Indices $l_{\gamma,k}$ 
for $\lambda_{\gamma,k} > \lambda_{\alpha,i}$ and $l_0$ stay 
unchanged:  $l'_{\gamma,k} = l_{\gamma,k}$, $l'_0=l_0$.  In the 
case if parameter $z$ crosses the cut on the right from the 
three--body threshold $\lambda_0$ (at $E>\lambda_0$) then all 
the indices $l_{\gamma,k}$ change as was described above.  
Besides, the index $l_0$ changes by 1, too. If at that, $z$ 
crosses the cut from below up then $l'_0=l_0+1$.  Otherwise 
$l'_0=l_0-1$.  Further, by $l$ we denote the multi--index 
$l=(l_0,l_1,l_2,l_3)$.

Thus, we have described the Riemann surface of the auxiliary 
vector--function $\sff(z)$.

As mentioned above we shall consider only a part of the 
three--body Riemann surface which will be denoted by $\Re$.  We 
include in $\Re$ all the sheets $\Pi_l$ of the Riemann surface 
of the function $\sff(z)$ with $l_0=0$. Also, we include in 
$\Re$ the upper half--plane, $\Img z > 0$, of the sheet $\Pi_l$ 
with $l_0=+1$ and the lower half--plane, $\Img z < 0$ , of the 
sheet $\Pi_l$ with $l_0=-1$.  For these parts we keep the 
previous notations  $\Pi_l$, $l_0=\pm 1$, assuming additionally 
that cuts are made on them along the rays belonging to the set 
$Z_{\rm res}=\Bigcup_{\an=1}^3 Z_{\rm res}\au$.  Here, 
$
Z_{\rm res}\au=
\{z:\, z=z_r \rho, \, 1\leq\rho <+\infty,\,
z_r\in\sigma\au_{\rm res} \}
$
is a totality of the rays beginning at the resonance points 
$z_r\in\sigma\au_{\rm res}$ of the subsystem $\an$ and going to 
infinity along the directions  $\hat{z}_r ={z_r}/{|z_{r}|}$.

The sheet $\Pi_l$ for which all the components of the 
multi--index  $l$ are zero, $l_0=l\ajd=0,$ $\anum$, is called 
physical sheet. The unphysical sheets $\Pi_l$ with $l_0=0$ are 
called two--body sheets since these ones may be reached rounding 
the two--body thresholds only and it is not necessary to round 
the three--body threshold $\lambda_0$.  The sheets $\Pi_l$ at 
$l_0=\pm 1$ are called three--body sheets. 

On the base of Sec.~\ref{SMSphys} results one can prove the following 
\begin{lemma}\label{LPossibility}\hspace*{-0.5em}{\sc .}
For each two--body unphysical sheet $\Pi_l$ of the surface 
$\Re$ there exists such a path from the physical sheet $\Pi_0$ 
to the domain $\Pi_l^{\rm (hol)}$ of $\Pi_l$ going only 
on the two--body unphysical sheets $\Pi_{l'}$ that moving by this path,  
the parameter $z$ stays always in respective domains 
$\Pi_{l'}^{\rm (hol)}\subset\Pi_{l'}$.
\end{lemma}

%
\section{\hspace*{-1em}. 
                   CONTINUATION OF THE FADDEEV EQUATIONS AND 
\newline
                   REPRESENTATIONS FOR MATRIX $M(z)$, SCATTERING MATRICES 
\newline
                   AND RESOLVENT ON UNPHYSICAL SHEETS }
\label{SRepresent}
In the present section we formulate main results  of the paper. 
In view of space shortage their proofs will be given 
in the following paper~\cite{Mot3Unphys2}. Here we outline only schemes 
of the proofs. 

We begin with description of continuation on unphysical sheets 
of the Faddeev equations~(\ref{Fadin}).

Let $L\au=$ $\diag\{ l_{\an,1},$ $l_{\an,2},...,$ $l_{\an,n\ad} \}$
be the diagonal number matrix constructed of the components 
$l_{\an,1},$ $l_{\an,2},...,$
$l_{\an,n\ad}$ of the multi--index $l$ identifying a certain 
sheet $\Pi_l\subset\Re$. 
For all this $L_1(l)=\diag\{ L^{(1)},$ $L^{(2)},$ $L^{(3)} \}$ and 
$L(l)=\diag\{L_0,$ $L_1\}$ á $L_0\equiv l_0$.

Let $\bs_{\an,l}(z)$ be the operator defined in $\hat{{\cal H}}_0$ by 
\be
\label{spair6}
\bs_{\an,l}(z)=\hat{I}_0+\rJo(z)\bt\ad(z)\rJot(z)A_0(z)L_0,
\quad z\in\Pi_0.
\ee
It follows from Eq.~(\ref{spair6}) that $\bs_{\an,l}=\hat{I}_0$ at $l_0=0$.
If  $l_0=\pm 1$ then according to Eqs.~(\ref{JTJS})--(\ref{J0t2J0t}), 
the operator $\bs_{\an,l}(z)$ is defined for $z\in\cP_b\bigcap\bC^\pm$
and acts on $f\in \hat{\cH}_0$ as 
\be
\label{spairh}
(\bs\adl(z)f)(\hP)=
\Int_{S^2} d\hk' s\ad(\hk\ad,\hk'\ad,z\cos^2\omega)
f(\cos\omega\ad\hk\ad',\sin\omega\ad\hp\ad),
\ee
where 
$
\hP=\lbrace \cos\omega\ad\hk\ad,\,\,\sin\omega\ad\hp\ad\rbrace
$
and  $s\ad$ is the scattering matrix~($\ref{s2}$) for the pair 
subsystem $\alpha$.  We take into account here the fact that 
$l_0\cdot{\rm Sign}\,\Img z=1$ for $l_0=1$ as well as $l_0=-1$.  
Remember that for $l_0=1$\,\, the sheet $\Pi_l$ is actually the 
upper half--plane $\bC^+$ and for $l_0=-1$, the lower one, 
$\bC^+$ (in accordance with our choice in 
Sec.~\ref{SRiemannSurface} of the part $\Re$ of the total 
three--body Riemann surface).  Therefore one can see now that on 
the both three--body sheets $\Pi_l$, $l_0=\pm 1$,  the operators 
$\bs_{\an,l}$ are described by the same formula (\ref{spairh}).  
As a matter of fact, the operators $\bs\adl(z)$ represent the 
scattering matrix (\ref{s2}) for the pair subsystem $\an$ 
rewritten in the three--body momentum space.

It follows immediately from Eq.~(\ref{spairh}) that if 
$z\in\cP_b\bigcap\bC^\pm\setminus {Z}\au_{\rm res}$ then there 
exists the bounded inverse operator $\bs\adl^{-1}(z)$,\,\,
\newline
$
\bigl(\bs\adl^{-1}(z)f\bigr)(\hP)=
\Int_{S^2} d\hk' s\ad^{-1}(\hk\ad,\hk'\ad,z\cos^2\omega\ad)
f(\cos\omega\ad\hk'\ad,\sin\omega\ad\hp\ad)
$
where $s\ad^{-1}(\hk,\hk',\zeta)$ stands for the kernel of the 
inverse pair scattering matrix $s\ad(\zeta)$.

The operator $\bs\adl^{-1}(z)$ becomes unbounded one at the 
boundary points $z$ belonging to  rims of the cuts 
(``resonance'' rays) included in $Z\au_{\rm res}$.

\begin{theorem}\label{ThKernels}\hspace*{-0.5em}{\sc .}
The absolute terms $\bt\ad(P,P',z)$ and kernels $(\bt\ad R_0)(P,P',z)$
of the Faddeev equations (\ref{Fadin})
admit the analytical continuation in a sense of distributions over $\Osix$
both on two--body  and three--body unphysical sheets $\Pi_l$
of the surface $\Re$.
The continuation on the sheet $\Pi_l$ 
with $l=(l_0,l_{1,1},...,$ $l_{1,n_1},l_{2,1},...,$
$l_{2,n_2},l_{3,1},...,l_{3,n_3})$, \,\,\,
 $l_0=0$, $l\bjd=0, 1,$
or $l_0=\pm 1$,  $l\bjd=1$
(in both cases $\bnum$)
read as 
\be
\label{tlRbig}
\bt^{l}\ad(z)\equiv \reduction{\bt\ad(z)}{\Pi_l}=
\bt\ad-L_0 A_0\bt\ad\rJot\bs\adl^{-1}\rJo\bt\ad-
\Phi\ad\rJ^{(\an)t}L\au A\au \rJ\au \Phi\ad^{*},
\ee
\be
\label{tr0l}
\reduction{[\bt\ad(z)R_0(z)]}{\Pi_l}=
\bt\ad^l (z)R_0^l(z)
\ee
where 
$
R_0^l(z)\equiv \reduction{R_0(z)}{\Pi_l}=
R_0(z)+L_0 A_0(z)\rJot(z)\rJo(z)
$
is the continuation~\cite{MotTMF} on $\Pi_l$ of the free Green 
function $R_0(z)$. If $l_0=0$ (and consequently, $\Pi_l$ is a 
two--body sheet) then the continuation (\ref{tlRbig}), 
(\ref{tr0l}) can be made on the hole sheet $\Pi_l$. For $l_0=\pm 
1$ (i.e.  in the case if $\Pi_l$ is a three--body sheet) the 
form (\ref{tlRbig}), (\ref{tr0l}) continuation is possible only 
on the domain  $\cP_b\bigcap\Pi_l$.  Al the kernels in r.h. 
parts of Eqs.~(\ref{tlRbig}) are taken on the physical sheet.
\end{theorem}

Proof of the theorem is based on utilizing the properties of the 
Cauchy type integrals (see Lemma from Sec.~2 of 
Ref.~\cite{MotTMF}), which are the integral terms of 
Eqs.~(\ref{Fadin}).

Using Eqs.~(\ref{tlRbig}) and~(\ref{tr0l}) one can rewrite the 
Faddeev equations (\ref{MFE}) continued on the sheet $\Pi_l$ in 
the matrix form  
\be
\label{MFEl}
M^l(z)=\bt^l(z)-\bt^l(z)\bRo^l(z)\Y M^l(z)
\ee
with 
\be
\label{tltot}
\bt^l(z)=\bt-
L_0 A_0 \bt\bJot\bs_l^{-1}\bJo\bt-\Phi\rJt_1 L_1 A_1\rJ_1\Phi^{*},
\ee
\be
\label{R0ltot}
\bRo^l(z)=\bRo(z)+L_0 A_0(z)\bJot(z)\bJo(z).
\ee
Here, 
$\bs_l(z)=\diag\{\bs_{1,l}(z),\bs_{2,l}(z),\bs_{3,l}(z)\}$.  By 
$M^l(z)$ we denote a supposed analytical continuation of the 
matrix $M(z)$ on the sheet $\Pi_l$.

\begin{theorem}\label{ThIter}\hspace*{-0.5em}{\sc .}
The kernels of the iterations $\cQ\un(z)=\bigl((-\bt\bRo\Y)^n 
\bt \bigr)(z)$,\,\, $n\geq 1$, allow, in a sense of 
distributions over $\Osix$, the analytical continuation on the 
domain $\Pilh$ of each unphysical sheet $\Pi_l\subset\Re$.  The 
continuation is described by 
$
%
  \reduction{\cQ\un(z)}{\Pi_l}=\bigl((-\bt^l\bRo^l\Y)^n \bt^l \bigr)(z).
$
\end{theorem}
\begin{note}\label{NIterF}\hspace*{-0.5em}{\sc .}
{\rm The products \,\,\, $L_1\rJ_1\Psis\Y\cQ^{(m)}$, \,\,\,\,\, 
$\cQ^{(m)}\Y\Psi\rJt_1 L_1$, \,\,\,\,\, 
$\tilde{L}_0\bJo\cQ^{(m)}$, \,\,\,\,\, 
$\cQ^{(m)}\bJot\tilde{L}_0$, \,\,\,\,\,\, 
\newline
$L_1\rJ_1\Psis\Y \cQ^{(m)}\Y\Psi\rJt_1 L_1$, \,\,\,\,
$\tilde{L}_0\bJo\cQ^{(m)}\bJot\tilde{L}_0,$ \,\,\,\,
$L_1\rJ_1\Psis\Y\cQ^{(m)}\bJot\tilde{L}_0$\,\, and 
$\tilde{L}_0\bJo\cQ^{(m)}\Y\Psi\rJt_1 L_1$,\,\, $0\leq m < n$, 
arising at substitution of the relations (\ref{tltot})
and (\ref{R0ltot})  into  
$ \reduction{\cQ\un(z)}{\Pi_l}$, have to be understood  
in a sense of the definitions of Sec.~\ref{SMSphys}. }
\end{note}
\begin{note}\label{NDomainF}\hspace*{-0.5em}{\sc .}
{\rm Theorem~\ref{ThIter} means that one can 
pose the continued Faddeev equations  (\ref{MFEl}) only in domains 
$\Pilh\subset\Pi_l$.
}
\end{note}

Construction of the representations for $M^l(z)$ consists 
actually in an explicit ``solving'' the continued Faddeev 
equations~(\ref{MFEl}) in the same way as in 
Refs.~\cite{MotTMF}, \cite{MotYaF} where the type~(\ref{3tnf}) 
explicit representations had been found for analytical 
continuation of $T$--matrix on unphysical sheets of the energy 
Riemann surface in the multichannel scattering problem with 
binary  channels.  Utilizing the expressions~(\ref{tltot}) for 
$\bt^l(z)$ and~(\ref{R0ltot}) for $\bRo^l(z)$, we begin with 
transfer of all the summands including $M^l(z)$ without $\bJo$ 
and $\rJ_1$ to the left--hand part of~(\ref{MFEl}). Then [for 
$z\not\in\sigma(H)$] we inverse the operators 
$\bI+\bt(z)\bRo(z)\Y$, using the relation 
$(\bI+\bt\bRo\Y)^{-1}=$ $\bI-M\Y\bRo$ (see Ref.~\cite{MF}).  
Introducing the new unknowns
$$
\begin{array}{l}
 \Xol=|L_0| \bs_l^{-1}\bJo(\bI-\bt\bRo)\Y M^l,\\
 \Xil=-L_1 \left[ \rJ_1\Phis\bRo+A_0 L_0\rJ_1\Phis\bJot\bJo\right]\Y M^l,
\end{array}
$$
we obtain for them a closed system of equations which was 
succeeded to solve explicitly.  Expressing then $M^l(z)$ by 
$\Xol$ and $\Xil$ one comes to the desired representations for 
$M^l(z)$.
\begin{theorem}\label{ThMlRepr}\hspace*{-0.5em}{\sc .}
The matrix $M(z)$ admits in a sense of distributions over 
$\Osix$, the analytical continuation in $z$ on the domains 
$\Pilh$ of unphysical sheets $\Pi_l$ of the surface $\Re$.  The 
continuation is described by 
\be
\label{Ml3fin}
M^l=M-\left(M\Omt\rJot,\,\,\,\,   \Phi\rJt_1+M\Y\Psi\rJt_1\right)
LA\,\, S_l^{-1}  \tL  \left(
\begin{array}{c}
\rJo\Om M \\ \rJ_1\Psis\Y M+ \rJ_1\Phis
\end{array}         \right)
\ee
where $S_l(z)$ is the truncated scattering matrix (\ref{Slcut}),\,\,
$L=\diag\{l_0,l_{1,1},...,$
$l_{1,n_1},$ $l_{2,1},...,$
$l_{2,n_2},$ $l_{3,1},...,$ $l_{3,n_3}\}$
and $\tL=\diag\{|l_0|,$ $l_{1,1},...,$
$l_{1,n_1},$ $l_{2,1},...,$
$l_{2,n_2},$ $l_{3,1},...,$ $l_{3,n_3}\}$.
Kernels of all the operators in the right--hand part of Eq.~(\ref{Ml3fin})
are taken on the physical sheet. 
\end{theorem}

Note that $LA\,\, S_l^{-1}(z)  \tL = \tL 
[S^\dagger_l(z)]^{-1}\,\, AL$.  Thus, the 
relations~(\ref{Ml3fin}) may be rewritten also in terms of the 
scattering matrices $S^\dagger_l(z)$.  It is clear that these 
relations may be rewritten in terms of symmetrized (truncated) 
scattering matrices~\cite{MerkDiss}, too. 

The representations for continuation of the (truncated) 
scattering matrices $S_l(z)$,\,\, $S_l(z):$ 
$\hat{\cH}_0\oplus\hat{\cH}_1\rightarrow\hat{\cH}_0\oplus\hat{\cH}_1$ 
and $\St_l(z)$,\,\, $\St_l(z):$ 
$\hat{\cH}_0\oplus\hat{\cH}_1\rightarrow\hat{\cH}_0\oplus\hat{\cH}_1$, 
follow from the representations (\ref{Ml3fin}) for $M^l(z)$.
Before writing final formulae we make some remarks. 

First of all, we note that the function $A_0(z)$ is 
univalent. It looks as $A_0(z)=-\pi i z^2$ on all the sheets $\Pi_l$. 
At the same time after continuing from $\Pi_0$ on $\Pi_l$, 
the function $A\bjd(z)=-\pi i\sqrt{z-\lambda\bjd}$
keeps its form if only $l\bjd=0$.
If $l\bjd=1$ this function turns into $A'\bjd(z)=-A\bjd(z)$.
Analogous inversion takes (or does not take) place 
for the arguments $\hP$, $\hP'$, $\hp\ad$ and $\hp'\bd$
of kernels of the operators 
$\rJo \Om M\Omt\rJot$,\,\, $\rJo \Om M\Y\Psi\rJt_1$,\,\,
$\rJ_1\Psis\Y M\Omt\rJot$ and $\rJ_1\Psis(\Y\bv+\Y M\Y)\Psi\rJt_1$, too.
Remember that on the physical sheet $\Pi_0$, 
the action of $\rJo(z)$\,\,  ($\rJot(z)$) transforms 
$P\in\Rs$ in $\sqrt{z}\hP$ \,\, ($P'\in\Rs$ in $\sqrt{z}\hP'$). 
At the same time, 
$p\ad\in\Rt$ \, ($p'\bd\in\Rt$)\, turns under $\rJ_{\an,i}(z)$\,\,
($\rJt_{\bn,j}(z)$)\,\, into 
$\sqrt{z-\lambda_{\an,i}}\, \hp\ad$ \,\,
($\sqrt{z-\lambda_{\bn,j}}\, \hp'\bd$).
Therefore we introduce the operators 
$\cE(l)=\diag\{\cE_0,\,\cE_1\}$ 
where $\cE_0$ is the identity operator in $\hat{\cH}_0$ if 
$l_0=0$, and $\cE_0$, the inversion 
$(\cE_0 f)(\hP)=f(-\hP)$ if $l_0=\pm 1$. Analogously,
$\cE_1(l)=\diag\{ \cE_{1,1},...,\cE_{1,n_1};$ $\cE_{2,1},...,\cE_{2,n_2};$
$\cE_{3,1},...,\cE_{3,n_3}\}$ where 
$\cE\bjd$ is the identity operator in $\hat{\cH}^{(\bn,j)}$ if
$l\bjd=0$, and $\cE\bjd$, the inversion 
$(\cE\bjd f)(\hp\bd)=f(-\hp\bd)$ if $l\bjd=1$.
By  $\re_1(l)$ we denote the diagonal matrix 
$\re_1(l)=\diag\{
\re_{1,1},...,\re_{1,n_1};$ $\re_{2,1},...,\re_{2,n_2};$
$\re_{3,1},...,\re_{3,n_3}\}$
with the elements  $\re\bjd=1$ if $l\bjd=0$ and $\re\bjd=-1$ if $l\bjd=1$.
Let  $\re(l)=\diag\{ \re_0, \re_1 \}$
where $\re_0=+1$.
\begin{theorem}\label{ThS3lRepr}\hspace*{-0.5em}{\sc .}
If there exists a path on the surface $\Re$ such that 
at moving by it from the domain $\Pilh$
on $\Pi_0$ to the domain $\Pilh\bigcap\Pi^{\rm (hol)}_{ll'}$
on $\Pi_{l'}$, the parameter $z$ stays on 
intermediate sheets $\Pi_{l''}$ always in the domains 
$\Pilh\bigcap\Pi^{\rm (hol)}_{ll''}$,
then the truncated scattering matrices $S_l(z)$ and $\St_l(z)$
admit analytical continuation in $z$
on the domain $\Pilh\bigcap\Pi^{\rm (hol)}_{ll'}$ of the sheet $\Pi_{l'}$.
The continuation is described by 
\begin{eqnarray}
\reduction{S_l(z)}{\Pi_{l'}} & = & \cE(l')\left[ \hbI+
\tL \hat{\cT} L\,\, A \re(l') -
\tL \hat{\cT} L'\,\, A\, S^{-1}_{l'}\,\, \tL' \hat{\cT} L\,\, A\re(l')
\right] \cE(l'),
\label{Slfin}                               \\
\reduction{\St_l(z)}{\Pi_{l'}} & = & \cE(l')\left[ \hbI+
 \re(l') A \,\, L \hat{\cT} \tL  -
\re(l') A\,\, L \hat{\cT} \tL'\,\, 
[\St_{l'}]^{-1}\, A \,\, L' \hat{\cT} \tL
\right] \cE(l')
\label{Stlfin}
\end{eqnarray}
where $L'=\left\{l'_0,\, l'_{1,1},...,l'_{1,n_1},\right.$
 $l'_{2,1},...,l'_{2,n_2},$
 $\left. l'_{3,1},...,l'_{3,n_3}\right\}$ and 
$\tL'=\left\{|l'_0|,\, l'_{1,1},...,l'_{1,n_1},\right.$
 $l'_{2,1},...,l'_{2,n_2},$
 $\left. l'_{3,1},...,l'_{3,n_3}\right\}$.
\end{theorem}

As we have established, the kernels of all the operators present 
in the right--hand part of expression~(\ref{RMR}) for the 
resolvent $R(z)$ admit, in a sense of distributions over 
$\Osix$,  the analytical continuation on the domains $\Pilh$ of 
unphysical sheets $\Pi_l\subset\Re$.  Hence, the kernel 
$R(P,P',z)$ admits such representation, too.
\begin{theorem}\label{ThResolvl}\hspace*{-0.5em}{\sc .}
The analytical continuation, in a sense of distributions over 
$\Osix$, of the resolvent $R(z)$ on the domain $\Pilh$ of 
unphysical sheet $\Pi_l\subset\Re$ is described by the formula 
\begin{eqnarray}
\nonumber
\lefteqn{\reduction{R(z)}{\Pi_l}=R+} \\
\label{R3l}
 & & +\bigl( [I-RV]\rJot\, ,\,\,\,\, \Om[\bI-\bRo M\Y]\Psi\rJt_1 \bigr)
LA S_l^{-1} \tL \left(
\begin{array}{c}  \rJo[I-VR]   \\     \rJ_1 \Psis[\bI-\Y M\bRo]\Omt
\end{array} \right).
\end{eqnarray}
Kernels of all the operators present in the right--hand part 
of Eq.~(\ref{R3l}) are taken on the physical sheet.
\end{theorem}

Note that in their structure, the representations~(\ref{R3l}) 
are quite analogous to that for analytical continuation of the 
two--body resolvent~(\ref{res2}).  Proof of the 
expressions~(\ref{R3l}) are based on immediate using the 
representations (\ref{Ml3fin}) for $M^l(z)$.
\section{\hspace*{-1em}. Acknowledgements}
The author is grateful to Professors~S.Albeverio, 
V.B.Belyaev, K.A.Ma\-ka\-rov and S.L.Ya\-kov\-lev for 
fruitful discussions. Also, the author is indebted to 
Prof.~P.Exner for useful remark.

\end{document}